\documentclass[11pt,letterpaper]{article}
\usepackage{times}
\usepackage{amsmath,amsthm,amsfonts}

\usepackage{verbatim,amsmath,amssymb,amsthm}
\usepackage[dvips]{graphics}
\usepackage{epsfig}
\usepackage{color}

\newcommand{\ignore}[1]{}

\newtheorem{theorem}{Theorem}[section]
\newtheorem{lemm}[theorem]{Lemma}

\newtheorem{claim}{Claim}[section]

\newtheorem{example}{Example}[section]
\newtheorem{thm}{Theorem}[section]
\newtheorem{deff}{Definition}

\newtheorem{fact}{Fact}[section]

\newcommand{\QSZK}{\mbox{QSZK}}
\newcommand{\SZK}{\mbox{SZK}}

\newcommand{\PSPACE}{\mbox{PSPACE}}

\newcommand{\SD}{\mbox{SD}}
\newcommand{\QSD}{\mbox{QSD}}
\newcommand{\ED}{\mbox{ED}}
\newcommand{\QED}{\mbox{QED}}
\newcommand{\EA}{\mbox{EA}}
\newcommand{\QEA}{\mbox{QEA}}

\newcommand{\Sn}{\mathcal{S}_n}
\newcommand{\An}{\mathcal{A}_n}

\newcommand{\C}{\mathbb{C}}

\newcommand{\F}{\mathbb{{F}}}

\newcommand{\set}[1]{{\left\{#1\right\}}}
\newcommand{\B}{\set{0,1}}

\newcommand{\Tr}{\mathop{\rm Tr}\nolimits}
\newcommand{\Span}{\mathop{\rm Span}\nolimits}

\newcommand{\perb}{\bot}
\newcommand{\Col}{\mathop{\rm Col}\nolimits}

\newcommand{\logeps}{\log({1 \over \epsilon})}
\newcommand{\poly}{\mbox{poly}}

\newcommand{\minentropy}{H_{\infty}}

\newcommand{\lbar}{\ol{\lambda}}

\newcommand {\ra} {\rangle}
\newcommand {\la} {\langle}
\newcommand{\ket}[1]{\left|#1\right\rangle}
\newcommand{\bra}[1]{\left\la #1 \right|}
\newcommand{\braket}[2]{\left\langle #1\!\mid\! #2\right\rangle}
\newcommand{\ketbra}[2]{\ket{#1}\!\bra{#2}}

\newcommand{\tensor}{\otimes}

\newcommand{\GL}{\mbox{GL}}
\newcommand{\PGL}{\mbox{PGL}(2,q)}

\newcommand{\CG}{\mathbb{C}[G]}

\newcommand{\CGG}{\mathbb{C}[G \times G]}

\newcommand{\norm}[1]{\left\|\,#1\,\right\|}

\newcommand{\trnorm}[1]{\norm{#1}_{\rm tr}}
\newcommand{\trn}[1]{\trnorm{#1}}

\newcommand{\nI}{\tilde{I}}

\newcommand{\res}{\mathrm{Res}}

\newcommand{\reg}{\rho_\mathrm{reg}}

\newcommand{\remove}[1]{}

\newcommand{\ol}[1]{\overline{#1}}

\newcommand{\half}{{1 \over 2}}

\newcommand {\wh} {\widehat}

\newsavebox{\savepar}

\newenvironment{boxit*}[0]
 { \begin{center}\begin{lrbox}{\savepar} \begin{minipage}[t]{\textwidth}}
 { \end{minipage}\end{lrbox}\fbox{\usebox{\savepar}} \end{center}}

\newenvironment{boxit}[0]
 { \begin{center}\begin{lrbox}{\savepar} \begin{minipage}[t]{\textwidth}}
 { \end{minipage}\end{lrbox}\fbox{\usebox{\savepar}} \end{center}}

\usepackage{latexsym}
\begin{document}

\newenvironment{Proof}{{\bf Proof:\ }}{\hfill$\Box$\medskip}
\newcommand{\pw}[1]{{\lfloor\!\!\lfloor#1\rfloor\!\!\rfloor}}

\newcommand{\coset}{\text{coset}}
\newcommand{\idx}{\text{index}}
\newcommand{\rank}{\mbox{\rm rank}}

\title{Quantum expanders and the quantum entropy difference problem}

\author{
\em Avraham Ben-Aroya
\thanks{School of Computer Science, Tel Aviv University, Tel Aviv
69978,Israel. E--mail: \{abrhambe,amnon\}@tau.ac.il} \and \em Amnon
Ta-Shma ${}^*$ }

\date{}

\maketitle

\begin{abstract}\noindent
We define quantum expanders in a natural way. We show that under
certain conditions classical expander constructions generalize to
the quantum setting, and in particular so does the Lubotzky, Philips
and Sarnak construction of Ramanujan expanders from Cayley graphs of
the group $\PGL$. We show that this definition is exactly what is
needed for characterizing the complexity of estimating quantum
entropies.


\end{abstract}

%
\section{Introduction}\label{sec:intro}

Expanders can be defined either combinatorially or algebraically. In
the combinatorial definition a graph $G=(V,E)$ is a
$(K_{max},c)$-expander, if every set $A \subseteq V$ of cardinality
at most $K_{max}$ has at least $c|A|$ neighbors. In the algebraic
definition we view $G$ as an operator defined by the normalized
adjacency matrix of the graph, and we say $G$ is a $\lbar$-expander
if the spectral gap between the first and second largest eigenvalues
(in absolute value) of this operator is at least $1-\lbar$.

We are interested in a sequence of graphs $\set{G_n}$, with an
increasing number of vertices, but constant degree $D$. The best
possible combinatorial expansion such a family can have is about
$D-2$, and the best possible algebraic expansion is about ${2
\over \sqrt{D}}$ (see \cite{N91}). The algebraic and combinatorial
definitions are closely related. Expanders with constant spectral
gap have constant combinatorial expansion and vice versa
\cite{AM85,A86}. However, this equivalence is not tight, and, in
particular, graphs with maximal spectral gap may have
combinatorial expansion not more than half the degree \cite{K95},
and graphs with almost optimal combinatorial expansion (close to
the degree) may have non-optimal spectral gap.

Both notions have proven extremely useful in computer science and
elsewhere. Often, the spectral gap is used (e.g., whenever a random
walk on the expander is used),  but sometimes combinatorial
expansion close to the degree is needed (e.g., in the  error
correcting codes of \cite{SS96}).

Thus, for decades, a major goal of computer science has been
constructing these marvelous graphs \emph{explicitly}.  Pinsker
\cite{P73} was the first to observe that \emph{non-explicitly},
constant degree expanders with very good combinatorial expansion
exist. Several explicit constructions of constant degree, algebraic
expanders with some constant (bounded away from zero) spectral gap
were given, e.g., in \cite{M73,GG81,JM87}. Lubotzky, Philips and
Sarnak \cite{LPS88} and Margulis \cite{M88} gave the first Ramanujan
graphs, i.e., a family $\set{G_n}$ of degree $D$ graphs with $\lbar$
approaching the optimal value. All the above graphs are Cayley
graphs and their analysis is algebraic. More recently, \cite{RVW00}
gave a more combinatorial construction, that was used in
\cite{CRVW02} to construct an explicit construction of graphs with
almost optimal \emph{combinatorial} expansion.

We refer the interested reader to the excellent survey paper
\cite{HLW06} for a comprehensive treatment of expander graphs, their
construction and applications.

\subsection{Quantum expanders}

Expanders are often thought of as \emph{combinatorial} objects. In
this view, expanders are sparse graphs that have combinatorial
expansion properties similar to random graphs. It is difficult to
see in this view how to generalize the notion to the quantum world.

However, most expander constructions, and many of the applications
that use expanders, treat expanders as algebraic objects, i.e., the
graph $G=(V,E)$ is translated to a linear mapping ${\cal G}$ from
some vector space ${\cal V}$ to itself. Let us describe how this is
done. Say $G=(V,E)$ is a graph. We translate $V$ to a vector space
${\cal V}$ of dimension $|V|$, with a basis vector $\ket{v}$ for
each $v \in V$. A probability distribution over $V$ then translates
to a vector $\sum_{v} p_v \ket{v}$ in this space, with $0 \le p_v
\le 1$ and $\sum_v p_v=1$. The graph $G$ is translated to the linear
operator ${\cal G}$ from ${\cal V}$ to ${\cal V}$ which is defined
by the normalized adjacency matrix of $G$. ${\cal G}$ is therefore a
linear mapping  ${\cal G}: {\cal V} \to {\cal V}$ that can be
classically implemented, and maps probability distributions to
probability distributions.




We extend the algebraic definition to the quantum setting. A general
classical state is a classical probability distribution over the
standard basis $\set{\ket{v}}$ of ${\cal V}$, i.e., vectors of the
form $\sum_v p_v \ket{v}$ as above. A general quantum state is a
\emph{density matrix} $\rho=\sum p_v \ketbra{\psi_v}{\psi_v}$, with
$0 \le p_v \le 1$, $\sum p_v=1$ and $\set{\psi_v}$ being \emph{some}
orthonormal basis of ${\cal V}$. In the classical world we had a
linear operator ${\cal G}: {\cal V} \to {\cal V}$. In the quantum
world a feasible quantum state is a matrix over ${\cal V}$, i.e., an
element of $L({\cal V})$, where $L({\cal V})$ is the set of linear
operators (matrices) over ${\cal V}$. We look for a linear
transformation $E: L({\cal V}) \to L({\cal V})$. Such a
transformation is called a \emph{super-operator}. We want in
addition that $E$ can be implemented by some physical process, and
this also ensures that $E$ maps density matrices to density
matrices. Such a linear operators $E$ is called in the literature an
\emph{admissible} super-operators.

We now turn to the regularity condition. Any (directed or
undirected) $D$-regular graph $G$ can have its edges labeled with
$1$ to $D$ such that each label $d \in [D]$ defines a
\emph{permutation} mapping. We define:

\begin{deff}
\label{def:qunatum-degree} We say an admissible super-operator $E:
L({\cal V}) \to L({\cal V})$ is \emph{$D$-regular} if $E={1 \over D}
\sum_d E_d$, and for each $d \in [D]$, $E_d(X)=U_d X U_d^\dagger$
for some unitary transformation $U_d$ over ${\cal V}$.
\end{deff}

In fact, for many classical constructions the edge labeling is
explicitly described in the construction, and in particular this is
always true whenever $G$ is a Cayley graph. This property was also
exploited in several constructions (e.g., in \cite{RVW00,CRVW02}).

Intuitively, a quantum expander is an admissible super-operator $E$
that has a spectral gap. We normalize the operator $E$ so that its
largest eigenvalue is $1$. As in the classical case we want the
eigenvector of eigenvalue one to be the completely mixed state. We
require that all other eigenvalues have a much smaller absolute
value. In general, however, $E$ need not be normal. This already
happens in the classical setting whenever we deal with directed
graphs. In such a case we need to replace eigenvalues with singular
values. Equivalently, we define:

\begin{deff}
\label{def:qexpander} An admissible superoperator $E:L(V) \to L(V)$
is a $(D,\lbar)$ expander if $E$ is $D$-regular and:

\begin{itemize}
\item
$E(\nI)=\nI$ and the eigenspace of eigenvalue $1$ has dimension $1$.

\item
For any $A \in L(V)$ that is orthogonal to $\nI$ (with respect to
the Hilbert-Schmidt inner product, i.e. $\Tr(A \nI) = 0$) it holds
that $\norm{E(A)}_2 \le \lbar \norm{A}_2$.
%
\end{itemize}
A quantum expander is \emph{explicit} if $E$ can be implemented by a
polynomial size circuit.
\end{deff}

Equivalently, we could have replaced the second condition with the
requirement  that all singular values of $T$ other than the largest
one (which is $1$) are smaller than $\lbar$.

\subsection{Are there any non-trivial quantum expanders?}

This is indeed a good question, and a major goal of this paper. A
first natural attempt is converting a good classical Cayley
expander, to a quantum super-operator. This indeed can be done, and
the resulting super operator $T: L({\cal V}) \to L({\cal V})$ is
analyzed in Section \ref{sec:T}. The analysis there shows that $T$
has $|V|$ eigenspaces, each of dimension $|V|$, with eigenvalues
$\overrightarrow{\lambda}=(\lambda_1=1,\ldots,\lambda_{|V|})$, where
$\overrightarrow{\lambda}$ is the spectrum of the Cayley graph. In
particular, the eigenspace of eigenvalue $1$ has dimension $|V|$
instead of dimension $1$.

Never the less, Ambainis and Smith obtained the following quantum
expander that is implicit in their work:

\begin{thm}
\label{thm:AS} \cite{AS04} There exists an explicit $({ \log^2 N
\over \lbar^2},\lbar)$ quantum expander $E:L(V) \to L(V)$, where
$N=dim(V)$.
\end{thm}

Their quantum expander is based on the classical Cayley expander
over the Abelian group $\mathbb{Z}_2^n$. As explained before, taking
the quantum analogue of the classical expander is not enough, and
Ambainis and Smith obtain their result using a clever trick,
essentially working over $\F_4^n$ rather than $\mathbb{Z}_2^n$.

The main problem with Abelian groups is that it is impossible to get
a constant degree Cayley expander over them \cite{K84,AR94}. This is
reflected in the $O(\log N)$ term in Theorem \ref{thm:AS}. There are
constant degree, Ramanujan Cayley graphs, i.e., Cayley graphs that
achieve the best possible relationship between the degree and the
spectral gap, but they are built over non-Abelian groups. If one
wants to get a constant degree quantum expander, then he is forced
to work over non-Abelian groups. Can one get constant degree quantum
expanders at all?

Our main construction starts with the constant degree Ramanujan
expander of \cite{LPS88}. This expander is a Cayley graph over the
non-Abelian group $\PGL$. We prove:

\begin{thm}\label{thm:PGL-expander}
There exists a $(D=O({1 \over \lbar^4}),\lbar)$ quantum expander.
\end{thm}

Our construction is not explicit in the sense that it uses the
Fourier transform over \PGL, which is not known to have an efficient
implementation (see \cite{LR92} for a non-trivial, but still not
fast enough, algorithm).

The \PGL{} quantum expander is as follows: we take two steps on
the classical expander graph, with a basis change between the two
steps. The basis change is a carefully chosen refinement of the
Fourier transformation that maps the standard basis $\ket{g}$
to the basis of the irreducible, invariant subspaces of $\PGL$.
Intuitively, in the Abelian case this basis change corresponds to
dealing with both the bit and the phase levels, and is similar to
the construction of quantum error correcting codes by first applying
a classical code in the standard basis and then in the Fourier
basis. However, this intuition is not as clear in the non-Abelian
case. Furthermore, in the non-Abelian case not every Fourier
transform is good. In this work we single out a natural algebraic
property we need from the underlying group that is sufficient for
proving the spectral gap of the construction. We then prove that
$\PGL$ respects this property.

We mention that there are also explicit, constant degree
(non-Ramanujan) Cayley expanders over $\Sn$ and $\An$ \cite{K05}.
Also, there is an efficient implementation of the Fourier transform
over $\Sn$ \cite{B97}. We do not know, however, whether $\Sn$ (or
$\An$) respect our additional property. We discuss this in more
detail in Section \ref{sec:Sn-construction}.

To summarize, Ambainis and Smith showed that good
\emph{poly-logarithmic-degree} quantum expanders exist, and their
construction is \emph{explicit}. Theorem \ref{thm:PGL-expander}
shows that good \emph{constant} degree quantum expanders
\emph{non-explicitly} exist (with a degree that is the square of
the degree of a Ramanujan graph). Recently, we showed together
with Oded Schwartz \cite{BST07} that one can use Theorem
\ref{thm:PGL-expander} with a Zig-Zag like construction, to obtain
an \emph{explicit}, \emph{constant} degree quantum expander.

Finally, we show a lower bound on the best achievable spectral
gap of quantum expanders.
\begin{thm} \label{thm:qexpander-lower-bound}
Any $(D,\lbar)$ quantum expander satisfies $\lbar \ge \frac{2}{3
\sqrt{3D}}$.
\end{thm}
The lower bound differs by a constant from the tight lower bound
known on classical expanders.

\subsection{What are quantum expanders good for?}

The first application of quantum expanders was given by Ambainis and
Smith themselves. They used these expanders to construct short
quantum one-time pads. Loosely speaking, they showed how two parties
sharing a random bit string of length $n + O(\log n)$ can
communicate an $n$ qubit state such that any eavesdropper cannot
learn much about the transmitted state. (A subsequent work by
\cite{DN06} showed how to remove the $O(\log n)$ term.)

In this paper we show another application of quantum expanders.
Watrous \cite{W02} defined the class of quantum statistical zero
knowledge languages (\QSZK). \QSZK{} is the class of all languages
that have a quantum interactive proof system, along with an
efficient simulator that produces transcripts that for inputs in the
language are statistically close to the correct ones (for the
precise details see \cite{W02,W06}).

Watrous defined the Quantum State Distinguishability promise problem
($QSD_{\alpha,\beta}$):

\begin{boxit}
\textbf{Input:} Quantum circuits $Q_0, Q_1$.

\textbf{Accept:} If $\trnorm{\ket{Q_0} - \ket{Q_1}} \ge \beta$.

\textbf{Reject:} If $\trnorm{\ket{Q_0} - \ket{Q_1}} \le \alpha$.
\end{boxit}

where the notation $\ket{Q}$ denotes the mixed state obtained by
running the quantum circuit $Q$ on the initial state $\ket{0^n}$
and tracing out the non-output qubits \footnote{Here we assume
that a quantum circuit also designates a set of output qubits.},
and $\trn{A}=\Tr{|A|}$ is the quantum analogue of the classical
$\ell_1$-norm (and so in particular $\trn{\rho_1-\rho_2}$ is the
quantum analogue of the classical variational distance of two
probability distributions).

Watrous showed $\QSD_{\alpha,\beta}$ is complete for
honest-verifier-$\QSZK$ ($\QSZK_{\text{HV}}$) when $0 \le \alpha <
\beta^2 \le 1$. He further showed that $\QSZK_{\text{HV}}$ is closed
under complement, that any problem in $\QSZK_{\text{HV}}$ has a $2$
message proof system and a $3$ message public-coin proof system and
also that $\QSZK \subseteq \PSPACE$. Subsequently, in \cite{W06}, he
showed that $\QSZK_{\text{HV}} = \QSZK$.

The above results have classical analogues. However, in the
classical setting there is another canonical complete problem, the
Entropy Difference problem ($\ED$). There is a natural quantum
analogue to \ED, the Quantum Entropy Difference problem (\QED), that
we now define:

\begin{boxit}
\textbf{Input:} Quantum circuits $Q_0, Q_1$.

\textbf{Accept:} If $S(\ket{Q_0}) -  S(\ket{Q_1}) \ge \half$.

\textbf{Reject:} If $S(\ket{Q_1}) -  S(\ket{Q_0}) \ge \half$.
\end{boxit}

where $S(\rho)$ is the Von-Neumann entropy of the mixed state
$\rho$.\footnote{A density matrix $\rho$ is positive semi-definite
and has trace $1$. Therefore its eigenvalues are all non-negative
and sum up to $1$, and can be thought of as defining a probability
distribution. The Von-Neumann entropy of $\rho$ is the Shannon
entropy of the eigenvalues of $\rho$.} We show that $\QED$ is
$\QSZK$-complete. We mention that for this purpose the expanders of
Ambainis and Smith given in Theorem \ref{thm:AS} suffice.

The problem $\QED$ is very natural from a physical point of view.
For example, a common way of measuring the amount of entanglement
between registers $A$ and $B$ in a pure state $\psi$ is by the
Von-Neumann entropy of $\Tr_B(\ketbra{\psi}{\psi})$ \cite{PR97}. Now
suppose we are given two circuits $Q_0$ and $Q_1$, both acting on
the same initial pure-state $\ket{0^n}$, and we want to know which
circuit produces more entanglement between $A$ and $B$. Our result
shows that this problem is $\QSZK$--complete. This, in particular,
shows that the harder problem of \emph{estimating} the amount of
entanglement between two registers in a given pure-state is
$\QSZK$--hard.

We believe these two applications are a good indication to the
usefulness of this notion. We expect that with time other
applications will be found.

Our proof that $\QED$ is $\QSZK$-complete uses a quantum variant of
classical balanced extractors. We explain this variant in Section
\ref{sec:extractors}. We show there that good balanced quantum
extractors exist. Surprisingly, we believe that unlike the classical
case, unbalanced quantum extractors do not exist.

\subsection{Summary and organization}

In classical computation there is a long line of research studying
"conductors": objects that manipulate their source entropy, using
few independent random bits. This research resulted in beautiful
constructions of expanders and extractors, and an amazing variety
of applications. We initiate the study of such "conductors"
manipulating the entropy of \emph{quantum} systems.

On the one-hand we show that expander-based constructions generalize
to the quantum setting (with effort, and not always, but at least in
some important cases). On the other hand, we believe all the huge
body of work relating classical extractors, condensers and such that
map a huge universe to a much smaller universe, is not likely to
have a quantum analogue (see Section
\ref{sec:extractors}). We think this study
deserves interest at its own right.

We also show two neat applications for quantum expanders. One, that
was already given in \cite{AS04} and a new one that we give here: we
characterize the complexity of approximating entropies. This proof
generalizes classical ideas, together with new technical work that
is needed for the quantum setting.

The paper is organized as follows. After the preliminaries (Section
\ref{sec:prel}), we give an intuitive exposition of our constant
degree expander, and the analysis, in Section
\ref{sec:main:quantum-expanders-from-non-abelian}. A complete
treatment is given in Section
\ref{sec:quantum-expanders-from-non-abelian} in the Appendix. In
Section \ref{sec:extractors} we discuss extractors, and discuss why
we believe \emph{unbalanced} quantum extractors are not useful. The
final section is devoted to proving the completeness of $\QED$ in
$\QSZK$. Here, again, we give an intuitive exposition in the main
text, with the formal details in the Appendix.

\section{Preliminaries}\label{sec:prel}

%


We first define the classical Renyi entropy. Let
$P=(p_1,\ldots,p_m)$ be a classical probability distribution.
%
The \emph{Shannon entropy} of $P$ is $H(P)=\sum_{i=1}^m p_i \lg
\frac{1}{p_i}$.
%
The \emph{min-entropy} of $P$ is $\minentropy (P)=\min_i \lg
\frac{1}{p_i}$.
%
The \emph{Renyi entropy} of $P$ is $H_2 (P)=\lg \frac{1}{\Col(P)}$,
where $\Col(P)=\sum p_i^2$ is the collision probability of the
distribution defined by $\Col(P)=\Pr_{x,y} [x=y]$ when $x,y$ are
sampled from $P$.

Now let $\rho \in D(V)$ be a density matrix (where $V$ is a Hilbert
space, $L(V)$ is the set of linear operators over $V$ and $D(V)$ is
the set of positive semi-definite operators in $L(V)$ with trace
$1$, i.e., all density matrices over $V$). Let
$\alpha=(\alpha_1,\ldots,\alpha_N)$ be the set of eigenvalues of
$\rho$. Since $\rho$ is positive semi-definite, all these
eigenvalues are non-negative. Since $\Tr(\rho)=1$ their sum is $1$.
Thus we can view $\alpha$ as a classical probability distribution.
%
The \emph{von Neumann entropy} of $\rho$ is $S(\rho)=H(\alpha)$.
%
The \emph{min-entropy} of $\rho$ is $\minentropy (\rho)=\minentropy
(\alpha)$.
%
The \emph{Renyi entropy} of $\rho$ is $H_2(\rho)=H_2(\alpha)$.
%
The analogue of the collision probability is simply $ \Tr({\rho^2})
= \sum_i \alpha_i^2 =||\rho||_2^2$. We remark that for any
distribution $P$, $\minentropy(P) \le H_2(P) \le H(P)$ and $2
\minentropy(P) \ge H_2(P)$.

%
%


The \emph{statistical difference} between two classical
distributions $P=(p_1,\ldots,p_m)$ and $Q=(q_1,\ldots,q_m)$ is
$\text{SD}(P,Q) = \half \sum_{i=1}^m |p_i - q_i|$, i.e., half the
$\ell_1$ norm of $P-Q$. This can be generalized to the quantum world
by defining the trace-norm of a matrix $X \in L(V)$ to be
$\trnorm{X}=\Tr(|X|)$, where $|X|=\sqrt{X X^{\dagger}}$,
and defining the \emph{trace distance} between density matrices
$\rho$ and $\sigma$ to be $\half \trnorm{\rho-\sigma}$.

%
%
%
%


\section{Quantum expanders from non-Abelian Cayley graphs}
\label{sec:main:quantum-expanders-from-non-abelian}

As we said before, our quantum expander takes two steps on a Cayley
expander (over the group \PGL) with a basis change between each of
the steps, and the basis change is a carefully chosen
transformation. In this section we give a bird's view of the proof.
We focus on the ideas, obstacles and solutions, and try to give an
informal presentation.

Our starting point is generalizing a single step on a Cayley graph
to the quantum setting. We fix an arbitrary (Abelian or non-Abelian)
group $G$ of order $N$, and a subset $\Gamma$ of group elements
closed under inverse. The \emph{Cayley graph} associated with
$\Gamma$, $C(G,\Gamma)$, is a graph over $N$ vertices, with an edge
between $(g_1,g_2)$ iff $g_1=g_2 \gamma$ for some $\gamma \in
\Gamma$. Rather then thinking of the Cayley graph as a graph, we
prefer to think of it as the linear operator over $\CG$ associated
with the adjacency matrix of $G$, where $\CG$ is the vector space
spanned by the basis elements $\ket{g}$ for each $g \in G$. I.e., it
is the linear operator $M=\frac{1}{|\Gamma|} \sum_{\gamma \in
\Gamma} \ketbra{x \gamma}{x}$.

We now define our basic superoperator $T: L(\CG) \to L(\CG)$. The
superoperator has a register $R$ of dimension $|\Gamma|$ that is
initialized at $\ket{\ol{0}}$. It does the following:

\begin{itemize}
\item
It first applies Hadamard on register $R$ (getting into the density
matrix ${1 \over |\Gamma|} \rho \tensor \sum_{\gamma,\gamma' \in
\Gamma} \ketbra{\gamma}{\gamma '}$).

\item
Then, it applies the unitary transformation $Z: \ket{g,\gamma} \to
\ket{g \gamma,\gamma}$. This transformation is a permutation over
the standard basis, and hence unitary. It is also classically easy
to compute in both directions, and therefore has an efficient
quantum circuit.

\item
Finally, it measures register $R$.
\end{itemize}

Thus we have: $T(\rho) =  \Tr_{R} [~Z (I \tensor H) (\rho \tensor
\ketbra{\ol{0}}{\ol{0}}) (I \tensor H) Z^\dagger ~]$. It can be
easily checked that over "classical" states (a density matrix $\rho$
that is diagonal in the standard basis) $T$ coincides with $M$.
Also, by definition, $T$ is $|\Gamma|$-regular.

The first thing to figure out is the eigenspace structure of the
super-operator $T$. This turns out to be as follows. $T$ has $N$
orthogonal eigen-spaces, each of dimension $N$, and the eigenvalues
$\lambda_1,\ldots,\lambda_N$ are those of $M$ (the orthogonality is
under the inner-product of $L(\CG)$ defined by $\la A | B \ra =
\Tr(A B^\dagger)$). In particular, if we start with a good Cayley
graph where $\lambda_1=1$ and all other eigenvalues have absolute
value at most $\lbar$, then $T$ has an eigenspace $W_1$ of dimension
$N$ with eigenvalue $1$, and all other eigenvalues have absolute
value at most $\lbar$. The fact that the dimension of $W_1$ is
larger than $1$ is not good for us, because it means that $T$ has no
spectral gap.

So, now we take a closer look at $W_1$ and we discover that it is
spanned by $\set{A_g~|~ g \in G}$ where $A_g=\sum_x \ketbra{gx}{x}$.
These operators $A_g$ are what is called the \emph{regular
representation} of $G$. Namely, if we denote $\reg(g)=A_g$, then
$\reg:G \to L(\CG)$ is a group homomorphism (namely, $\reg(g_1 \cdot
g_2)=\reg(g_1) \cdot \reg(g_2)$). Furthermore, a basic theorem of
representation theory says that there is a basis change under which
all the operators $A_g=\reg(g)$ simultaneously block-diagonalize,
with the blocks corresponding to the irreducible representations of
$G$. This (non-unique) basis change is called the Fourier transform
of $G$.

Let us first consider the case where $G$ is Abelian, and let $e$
denote the identity element in $G$. In this case all the irreducible
representations of $G$ have dimension one, and the Fourier transform
$U$ simultaneously diagonalizes all the operators $A_g=\reg(g)$. The
elements $\set{A_g=\reg(g)}$ form an orthonormal basis of $W_1$.
Doing the basis change, they all become diagonal, i.e., "classical"
states. Furthermore, $A_e=\reg(e)=I$ is mapped to $I$ (as is true in
any basis change) and all other basis elements are mapped to
orthogonal states (as $U$ is unitary). We therefore expect that
applying $T$ again now, is equivalent to applying $M$ on the
classical state, and will result in a unique eigenvector of
eigenvalue $1$, with all other eigenvalues being at most $\lbar$.

So our (Abelian) quantum expander is as follows. We let $U$ be the
Fourier transform over $G$, and the quantum expander is the
superoperator
$$E(\rho)= T( U T(\rho) U^\dagger).$$

A simple check shows that $E$ is indeed a $\lbar$--expander, and its
spectral gap is the same as that of $G$. Also, clearly, $E$ is
$|\Gamma|^2$-regular.

We now turn to the non-Abelian case. Here most irreducible
representations have dimension larger than $1$, and as a result the
basis change does not diagonalize all $A_g=\reg(g)$, but rather just
block-diagonalizes them, with blocks corresponding to the
irreducible representations. In particular, doing the Fourier
transform does not map $A_g=\reg(g)$ to "classical" states. Never
the less, this does not necessarily mean that the above approach
fails. In fact, it turns out that a sufficient requirement for a
good basis change is that for any $g_1 \ne e$ and any $g_2$, it
holds that

\begin{eqnarray}
\label{eqn:main:propertyU} \Tr(U \reg(g_1) U^\dagger \reg(g_2)) &=&
0.
\end{eqnarray}

Intuitively, we can do the analysis separately for elements in $W_1$
and elements in $W_1^{\perb}$ - the space perpendicular to $W_1$
(this is technically more complicated, see Lemma \ref{lem:col-sum}).
Elements in $W_1^\perb$ are immediately shortened by the first
application of $T$. Elements in $\text{Span} \set{A_g=\reg(g) ~|~ g
\neq e}$ are kept in place by the first application of $T$, but are
mapped to $W_1^\perb$ by the basis change, and therefore are
shortened by the second application of $T$. Together, if $U$ is a
good basis change then $E(\rho)= T( U T(\rho) U^\dagger)$ is a
$\lbar$--expander.

But does a good basis change always exist?

We consider the dihedral group as an illuminating example. The
dihedral group has irreducible representations of dimension $2$ (and
a few of dimension $1$). The dihedral group also has a cardinality
two subgroup $H=\set{e,s}$, where $s$ is the reflection element. The
Fourier transform associates the eigen-spaces of the irreducible
representations, to elements of $G$. Now, imagine that we associate
the dimension-$2$ blocks with cosets of $H$. A moment of thought
reveals that if $g_2 \not \in H$ then Equation
(\ref{eqn:main:propertyU}) is satisfied! This is because $A=U
\reg(g_1) U^\dagger$ has non-zero elements only on the $2$ by $2$
blocks, while $B=\reg(g_2)=\sum_x \ketbra{g_2 x}{x}$ has non-zero
elements only outside these $2$ by $2$ blocks, and so the inner
product $\la A | B \ra = \Tr(A B^\dagger) = \sum_{i,j} A_{i,j}
\ol{B_{i,j}}$ must be zero.

We need also to consider the case where $g_2 \in H=\set{e,s}$. If
$g_2=e$ then $\Tr(U \reg(g_1) U^\dagger \reg(g_2))=\Tr(\reg(g_1))$
and the analysis is simple. We are left with the case $g_2=s$.
Recall that $\Tr(A B^\dagger) = \sum_{i,j} A_{i,j} \ol{B_{i,j}}$. We
can interpret the expression $\Tr(U \reg(g_1) U^\dagger \reg(s))$ as
the sum of all entries $i,j$ of $U \reg(g_1) U^\dagger$ that belong
to the set $\text{P}=\set{(is,i)}$. We now use the fact that each
irreducible representation appears in the regular representation
with multiplicity that equals its dimension. In matrix language this
means that for each dimension $2$ irreducible representation, there
are two corresponding blocks in the decomposition, and the entries
in these two blocks can be made \emph{identical} (see Section
\ref{sec:rep-background} for more background on representation
theory). As the blocks correspond to cosets of $H$, multiplication
by $g_2=s$ has the same effect in the two cosets. I.e., an entry of
one block is in $\text{P}$ and is added to the sum, iff the
corresponding entry in the other block is also in $\text{P}$ and is
also added to the sum. We can therefore force a zero sum, by forcing
one block to be the negative of the other block, which can be done
by an easy manipulation of the Fourier transform.

At first, the above solution looks ad hoc, and very specific to the
dihedral group. So we try to abstract the ingredients that have been
used in the solution.

The Fourier transform is a unitary mapping from the standard basis
$\set{\ket{g}}$ of $\CG$, to the Fourier basis. It can be formally
defined as follows. Let $\widehat{G}$ denote the set of all
inequivalent irreducible representations of $G$. For a representation
$\rho$ let $d_{\rho}$ denote the dimension of $\rho$. We define the
transformation $F$ by

\begin{eqnarray*}
F \ket{g} &=& \sum_{\rho \in \widehat{G}} \sum_{1 \le i,j \le
d_\rho} \sqrt{\frac{d_\rho}{|G|}} \rho_{i,j}(g) \ket{\rho,i,j}.
\end{eqnarray*}

It can be checked that $F$ is unitary and that it indeed
block-digaonlizes the regular representations, namely,

\begin{eqnarray*} F \reg(g) F^\dagger &=&
   \sum_{\rho \in \widehat{G}} \sum_{1 \le j \le d_\rho}
   \ketbra{\rho,j}{\rho,j} \tensor \sum_{1 \le i,i' \le d_\rho} \rho_{i,i'}(g)
   \ketbra{i}{i}
\end{eqnarray*}

I.e., for each $\rho \in \wh{G}$ and $j \le d_\rho$, we have a
$d_\rho \times d_\rho$ block whose entries are $\rho(g)$.

$F$ maps $\CG$ to a vector space of the same dimension that is
spanned by  $\set{\ket{\rho,i,j} : \rho \in \widehat{G},~ 1 \le i,j
\le d_\rho}$. To complete the specification of the Fourier transform
we also need to specify a map $S$ between $\set{\ket{\rho,i,j}}$ and
$\set{\ket{g} : g \in G}$. In the Abelian case there is a canonical
map $S$ between $\set{\ket{\rho,i,j} : \rho \in \widehat{G},~ 1
=j=1}$ and $\set{\ket{g} : g \in G}$, because when $G$ is Abelian
$\widehat{G}$ is isomorphic to $G$. However, when $G$ is not Abelian
things are more complicated. It is always true that $\sum_{\rho \in
\widehat{G}} d_\rho^2 = |G|$, and so there is always a bijection
between $\set{\ket{\rho,i,j}}$ and $\set{\ket{g} : g \in G}$.
However, it is not known, in general, how to find such a natural
bijection.

For example, for the symmetric group $\Sn$ the question takes the
following form. We look for bijections $f$ from pairs $(P,T)$ of
standard shapes to $\Sn$ (a shape corresponds to an irreducible
representation of $\Sn$, and its dimension is the number of standard
shapes of that shape). The question of finding an explicit bijection
$f$ from pairs $(P,T)$ of standard shapes to $\Sn$ is a basic
question in the study of the representation theory of $\Sn$. The
canonical algorithm doing so is the "Robinson-Schensted" algorithm
\cite{R38,S61} that was extensively studied later on (see
\cite{S01}, and especially Chapter 3 that is almost completely
dedicated to this algorithm).

Looking back at the solution we gave for the dihedral group we see
that we can express it as follows. We made sure that a block that
corresponds to an irreducible representation is contained in a coset
of $H$, and different copies of the same representation get the same
indices within $H$. Generalizing this further, we see that what we
actually used is a mapping $S: \set{\rho,i,j} \to G$ that is
\emph{product}, i.e., for every $\rho \in \widehat{G}$,
$S(\rho,i,j)=f_1(i) \cdot f_2(j)$ for some functions $f_1,f_2:
[d_\rho] \times [d_\rho] \to G$ (the functions $f_1$ and $f_2$ may
be specific to $\rho$). In the dihedral group, this amounts to $f_2$
selecting a coset representative, and $f_1$ selecting an index
inside the coset. But, in fact, any product mapping $S$ is good.

It is not clear at all that for every group $G$ such a product
mapping exists. It is trivial for Abelian groups, and simple for the
dihedral group (using cosets of $\set{e,s}$ for example). It is not
clear what is the situation for $\Sn$ - the Robinson-Schensted is
not a product mapping, but using specific information about $\Sn$,
for $n \le 6$, we found out that a product mapping exists.
Never the less, we were able to prove that $\PGL$ has a product
mapping, using information about its subgroup structure, and its
irreducible representations.

Putting these things together, we get a quantum expander $E(\rho)=
T( U T(\rho) U^\dagger)$, with $T$ being a single quantum step on a
the Cayley expander, and $U$ being a good basis change. $U$ is
obtained by doing the standard Fourier transform $F$ followed by the
a product mapping $S$, and with adding appropriate phases to the
basis vectors, so as different copies of the same irreducible
representation cancel out.

Clearly, the above discussion is intuitive, and there are many gaps
to fill. This is done in Appendix
\ref{sec:quantum-expanders-from-non-abelian}, where we repeat
everything in a relaxed way and with all the necessary details. In
Sec \ref{sec:rep-background} we give some background on
representation theory. Section \ref{sec:T} analyzes a single quantum
step on a Cayley graph and in Section \ref{sec:Abelian-expander} we
analyze the quantum expander over Abelian groups. Section
\ref{sec:Template} singles  out Property (\ref{eqn:main:propertyU})
as a sufficient condition for a good basis change, and Section
\ref{sec:Property} shows that all we need for that is finding a
product mapping $S$. Finally, we prove in Section \ref{sec:PGL} that
$\PGL$ has such a product mapping, completing the correctness proof
of our constant degree quantum expander.


\section{Quantum extractors}
\label{sec:extractors}

\emph{\bf The balanced case.} The classical proof that $\ED$ is
$\SZK$-complete uses balanced \emph{extractors}. A balanced
extractor is a function $E: \B^n \times \B^d \to \B^n$. We say $E$
is a $(k,\epsilon)$ extractor if for every distribution $X$ on
$\B^n$ that has $k$ min-entropy
the
distribution $E(X,U_d)$ obtained by sampling $x \in X$, $y \in \B^d$
and outputting $E(x,y)$, is $\epsilon$--close to uniform. We now
define balanced quantum extractors.

\begin{deff}
\label{def:balanced-quantum-extractor} Let $V$ be a Hilbert space of
dimension $N$. A superoperator $T: L(V) \to L(V)$ is a
$(k,d,\epsilon)$ \emph{quantum extractor}, if $T$ is $2^d$-regular
and for every $\rho \in D(V)$ with
$\minentropy(\rho) \ge k$ we have $\trn{T\rho-\nI} \le \epsilon$,
where $\nI={1 \over N}I$. We say $T$ is efficient if $T$ can be
implemented by a polynomial-size quantum circuit.
\end{deff}

We
mention that if $T$ is $2^d$-regular (and, in particular, if it is a
$(k,d,\epsilon)$ quantum extractor) then for any $\rho \in L(V)$ it
holds that $S(T\rho) \le S(\rho)+d$, i.e., no matter what, the
extractor never adds more than $d$ entropy to any input system.

Classically, balanced extractors are closely related to expanders
(e.g., \cite{GW97}). This generalizes to the quantum setting. We
prove:

\begin{lemm}
\label{lem:expander-extractor} If $T:L(V) \to L(V)$ is a
$(D=2^d,\lbar)$ quantum expander, then for every $t>0$, $T$ is also
a $(k=n-t,d,\epsilon)$ quantum extractor with $\epsilon=2^{t/2}
\cdot \lbar$.
\end{lemm}

We give the easy proof in Section \ref{sec:app:quantum-extractors}
in the Appendix. In particular, we get an $(n-t,d,\epsilon)$
balanced quantum extractor $T:L(V) \to L(V)$ where $n=dim(V)$, and
$d=2(t + 2\logeps)+O(1)$ using Theorem \ref{thm:PGL-expander} (or
the explicit version given in \cite{BST07}).

We use the last lemma to prove our lower bound on the spectral
gap of quantum expanders.
\newtheorem*{thma}{Theorem~\ref{thm:qexpander-lower-bound}}
\begin{thma}
Any $(D,\lbar)$ quantum expander satisfies $\lbar \ge \frac{2}{3
\sqrt{3D}}$.
\end{thma}

In the classical world a tight bound of about ${2 \sqrt{D-1}} \over
D$ has been proved \cite{N91}. The proof there is both algebraic
(using eigenvalues) and combinatorial (using paths in the graph). We
do not see how to generalize the combinatorial component of the
proof. Instead we give an algebraic proof. The proof idea is to take
a density matrix which is uniform on a set of "small size". Applying
the extractor yields a density matrix close to the completely mixed
state. Such a matrix must have a high rank. On the other hand,
because we started with a low-rank matrix, the resulting density
matrix cannot have a too-high rank (since $E$ is $D$-regular). The
formal details are given in Section \ref{sec:app:quantum-extractors}
in the Appendix.

\emph{\bf The unbalanced case.}
%
%
%
A natural generalization of Definition
\ref{def:balanced-quantum-extractor} is for
%
a superoperator $T: L(V) \to L(W)$ where $V,W$ are Hilbert spaces of
dimensions arbitrary dimensions $N$ and $M$.
I.e., here we let $W$ be different than $V$, and, in particular, the
superoperator $T$ can map a large Hilbert space $V$ to a much
smaller Hilbert space $W$. In the classical case this corresponds to
hashing a large universe $\B^n$ to a much smaller universe $\B^m$.
Indeed, in the classical world highly unbalanced extractors exist
with a very short seed length $d$. These (and related objects like
dispersers, condensers and unbalanced expanders) have numerous
applications. There is also a huge body of work constructing
explicitly (most of) these objects. See \cite{CRVW02} for an attempt
to put some order in the zoo of definitions, and \cite{N96,S02} for
a survey of applications and constructions.

However, here we see a difference between the classical and the
quantum world. In the classical world if $X$ has $k$ entropy, and we
add $d$ more uniform bits, then the final output distribution can
have at most $k+d$ entropy. If we then "ignore" some of the output
bits, we can only \emph{decrease} the entropy of the output
distribution. In particular, if the output distribution has $m$
entropy, then most of it (namely, $m-d$) came from the source $X$.
We also had a similar property for balanced quantum extractors: for
any input $\rho$ we had $S(T \rho) \le S(\rho)+d$.

In the unbalanced case, however, we output $m \ll n$ qubits, and so
we trace-out (or "ignore") qubits. This, by itself, may
\emph{increase} the entropy. For example, a mixed state that is with
probability one in some pure-state has entropy zero (it is
completely determined). Tracing out $k$ bits of the system, may
result in a mixed state having $k$ entropy. If we trace out $n/2$
bits, at least theoretically, it is possible that our extractor
starts with a pure state as an input $\rho$ (i.e., $\rho$ has zero
entropy) and ends up with $T\rho$ being the completely mixed state.
Notice that at most $d$ of this entropy comes from the seed, and the
rest comes from the tracing-out.
%
%
%
We believe this makes any unbalanced extractor with $m<n/2$ not
useful. For example, the property $S(T\rho) \le S(\rho)+d$ (true for
balanced quantum extractors) is crucial for our proof that \QED{} is
\QSZK-complete.
%
We believe that slightly unbalanced expander
constructions (e.g., \cite{M95}) can probably be converted to
useful, slightly unbalanced quantum extractors.


\section{The complexity of estimating entropy}
\label{sec:main:estimating-entropy}

In this section we show that the $\QED$ problem (as defined in the
introduction) is $\QSZK$-complete. We do that by showing that $\QED$
reduces to $\QSD$ and vice versa, using the already known fact that
$\QSD$ is $\QSZK$--complete.

Proving $\QED{} \le \QSD$ is a bit tricky. We first show a that
related problem, Quantum Entropy Approximation ($\QEA$), reduces to
$\ol{\QSD}$. $\QEA$ is the following promise problem:

\begin{boxit}
\textbf{Input:} A Quantum circuit $Q$ and a non-negative integer $t$.

\textbf{Accept:} If $S(\ket{Q}) \ge t+\half$.

\textbf{Reject:} If $S(\ket{Q}) \le t-\half$.
\end{boxit}

$\QEA$ is the problem of comparing the entropy of a given quantum
circuit to some \emph{known} threshold $t$, instead of comparing the
entropies of two quantum circuits as in $\QED$. Our proof that $\QEA
\le \ol{\QSD}$ uses quantum expanders and extractors, and we discuss
it next.

We begin with the classical intuition why $\EA$ reduces to $\SD$
($\EA$ is the same promise problem, but with the input being a
classical circuit). We are given a circuit $C$ and we want to
distinguish between the cases the distribution it defines has
substantially more or less than $t$ entropy. First assume that the
distribution is flat, i.e., all elements that have a non-zero
probability in the distribution, have equal probability. In such a
case we can apply an extractor on the $n$ output bits of $C$,
hashing it to about $t$ bits. If the input distribution has
high entropy, it also has high min-entropy (because for flat
distributions entropy is the same as min-entropy) and therefore the
output of the extractor is close to uniform. If, on the other hand,
the circuit entropy is less than $t-d-1$, where $d$ is the extractor
seed length, than even after applying the extractor the output
distribution has at most $t-1/2$ entropy, and therefore it must be
far away from uniform. We get a reduction to $\ol{\SD}$.

There are, of course, a few gaps to complete. First, our source is
not necessarily flat. This is solved in the classical case by taking
many independent copies of the circuit, which makes the output
distribution "close" to "nearly-flat" . A simple analysis shows that
this flattening works also in the quantum setting. Also, we need to
amplify the gap we have between entropy $t+1/2$ and $t-1/2$ to a gap
larger than $d$ (the seed length). This, again, is solved by taking
many independent copies of $C$, because $S(C^{\tensor q})=q S(C)$,
and works the same way in the quantum setting.

The interesting question is what is needed in the quantum case from
the quantum analogue of classical extractors. As it turns out, what
is needed is that sources with high min-entropy are mapped close to
the completely mixed state, whereas \emph{all} sources of low
min-entropy are mapped far away from it. The first condition is
clearly satisfied by our Definition
\ref{def:balanced-quantum-extractor}. The second condition is
implied by the regularity of the extractor: a $D=2^d$ regular
extractor can never add more than $d$ entropy to a source, and so
sources with low min-entropy are mapped to sources with low
min-entropy, and such sources (with the right parameters) are far
away from uniform. The formal proof is given in Section
\ref{sec:QEA-QSD}.

We remark that we believe that exactly this property fails in the
unbalanced case, i.e., there are input sources with low min-entropy
(e.g. pure states) that are mapped close to the completely mixed
state, and this additional entropy is obtained not because of the
seed, but rather because we have an unbalanced extractor that traces
out registers.

This completes the proof that $\QEA$ reduces to $\ol{\QSD}$. As
Watrous showed that $\ol{\QSD} \le \QSD$, we get that $\QEA \le
{\QSD}$. We next show that $\QEA \le \QSD$ implies $\QED \le \QSD$
using a standard classical trick. We can express:
%
$\QED(Q_0,Q_1)  = \bigvee_{t=1} \left[((Q_0,t) \in \QEA_Y) \wedge
((Q_1,t) \in \QEA_N)\right]$.
%
Thus, if $\QEA$ reduces to $\QSD$ (as we proved), we can express
$\QED$ as a formula over $\QSD$. We then take the classical result
that any Boolean formula over $\SD$ reduces to $\SD$, and generalize
it to the quantum setting, concluding that $\QED$ reduces to $\QSD$
as desired. The full details (and this time just for completeness,
because the proof closely follows the classical one) are given in
Section \ref{sec:closure}. This completes the proof that $\QED \le
\QSD$.

The direction that $\QSD \le \QED$ follows the classical reduction,
but using the Holevo bound from quantum information theory. The
details are given in Section \ref{sec:QSD-QED}. Altogether, we see
that $\QED$ is $\QSZK$ complete.

\section*{Acknowledgements}

We thank Oded Regev for pointing out \cite{AS04} to us and for
referring us to Lemma \ref{lem:ANTV} that simplified the proof of
the reduction from $\QSD$ to $\QED$. We also thank Ashwin Nayak,
Oded Regev, Adam Smith and Umesh Vazirani for helpful discussions
about the paper.

\bibliographystyle{alpha}
\bibliography{bibl}

\begin{thebibliography}{CRVW02}

\bibitem[Alo86]{A86}
N.~Alon.
\newblock {Eigen values and expanders}.
\newblock {\em Combinatorica}, 6(2):83--96, 1986.

\bibitem[AM85]{AM85}
N.~Alon and V.~Milman.
\newblock {lambda sub (1), isoperimetric inequalities for graphs, and
  superconcentrators.}
\newblock {\em J. COMBINAT. THEORY SER. B.}, 38(1):73--88, 1985.

\bibitem[ANTV02]{ANTV02}
A.~Ambainis, A.~Nayak, A.~Ta{-}Shma, and U.~V. Vazirani.
\newblock Quantum dense coding and quantum finite automata.
\newblock {\em Journal of the ACM}, 49:496--511, 2002.
\newblock Earlier version in 31st ACM STOC, 1999, pp. 376-383.

\bibitem[AR94]{AR94}
N.~Alon and Y.~Roichman.
\newblock {Random Cayley Graphs and Expanders}.
\newblock {\em Random Structures and Algorithms}, 5(2):271--285, 1994.

\bibitem[AS04]{AS04}
A.~Ambainis and A.~Smith.
\newblock Small pseudo-random families of matrices: Derandomizing approximate
  quantum encryption.
\newblock In {\em RANDOM}, pages 249--260, 2004.

\bibitem[Bea97]{B97}
R.~Beals.
\newblock {Quantum computation of Fourier transforms over symmetric groups}.
\newblock {\em STOC}, pages 48--53, 1997.

\bibitem[BST07]{BST07}
A.~Ben{-}Aroya, O.~Schwartz, and A.~Ta{-}Shma.
\newblock An explicit, constant degree quantum expander.
\newblock unpublished manuscript, April 2007.

\bibitem[CRVW02]{CRVW02}
M.~Capalbo, O.~Reingold, S.~Vadhan, and A.~Wigderson.
\newblock Randomness conductors and constant-degree expansion beyond the degree
  / 2 barrier.
\newblock In {\em STOC}, pages 659--668, 2002.

\bibitem[DN06]{DN06}
P.~Dickinson and A.~Nayak.
\newblock Approximate randomization of quantum states with fewer bits of key.
\newblock In {\em AIP Conference Proceedings}, volume 864, pages 18--36, 2006.

\bibitem[GG81]{GG81}
O.~Gabber and Z.~Galil.
\newblock {Explicit Constructions of Linear-Sized Superconcentrators}.
\newblock {\em Journal of Computer and System Sciences}, 22(3):407--420, 1981.

\bibitem[GW97]{GW97}
O.~Goldreich and A.~Wigderson.
\newblock {Tiny families of functions with random properties: A quality-size
  trade-off for hashing}.
\newblock {\em Random Structures and Algorithms}, 11(4):315--343, 1997.

\bibitem[HF91]{FH91}
J.~Harris and W.~Fulton.
\newblock {\em Representation Theory}.
\newblock Springer, 1991.

\bibitem[HLW06]{HLW06}
S.~Hoory, N.~Linial, and A.~Wigderson.
\newblock Expander graphs and their applications.
\newblock {\em Bulletin of the AMS}, 43(4):439--561, 2006.

\bibitem[HRT00]{SRT02}
S.~Hallgren, A.~Russell, and A.~Ta{-}Shma.
\newblock Normal subgroup reconstruction and quantum computation using group
  representations.
\newblock In {\em STOC}, pages 627--635, 2000.

\bibitem[JM87]{JM87}
S.~Jimbo and A.~Maruoka.
\newblock Expanders obtained from affine transformations.
\newblock {\em Combinatorica}, 7(4):343--355, 1987.

\bibitem[Kah95]{K95}
N.~Kahale.
\newblock {Eigenvalues and expansion of regular graphs}.
\newblock {\em Journal of the ACM (JACM)}, 42(5):1091--1106, 1995.

\bibitem[Kas05]{K05}
M.~Kassabov.
\newblock Symmetric groups and expanders.
\newblock {\em Electron. Res. Announc. Amer. Math. Soc.}, 11, 2005.

\bibitem[Kla84]{K84}
M.~M. Klawe.
\newblock {Limitations on Explicit Constructions of Expanding Graphs}.
\newblock {\em SIAM J. Comput.}, 13(1):156--166, 1984.

\bibitem[LPS88]{LPS88}
A.~Lubotzky, R.~Philips, and P.~Sarnak.
\newblock Ramanujan graphs.
\newblock {\em Combinatorica}, 8:261--277, 1988.

\bibitem[LR92]{LR92}
J.~D. Lafferty and D.~Rockmore.
\newblock Fast fourier analysis for {${\rm SL}\sb 2$} over a finite field and
  related numerical experiments.
\newblock {\em Experiment. Math.}, 1(2):115--139, 1992.

\bibitem[Mar73]{M73}
G.~A. Margulis.
\newblock {Explicit constructions of expanders}.
\newblock {\em Problemy Peredaci Informacii}, 9(4):71--80, 1973.

\bibitem[Mar88]{M88}
G.~A. Margulis.
\newblock {Explicit group-theoretic constructions of combinatorial schemes and
  their applications in the construction of expanders and concentrators}.
\newblock {\em Problemy Peredachi Informatsii}, 24(1):51--60, 1988.

\bibitem[Mor95]{M95}
M.~Morgenstern.
\newblock {Natural bounded concentrators}.
\newblock {\em Combinatorica}, 15(1):111--122, 1995.

\bibitem[NC00]{NC00}
M.~Nielsen and I.~Chuang.
\newblock {\em Quantum Computation and Quantum Information}.
\newblock Cambridge University Press, 2000.

\bibitem[Nil91]{N91}
A.~Nilli.
\newblock On the second eigenvalue of a graph.
\newblock {\em Discrete Mathematics}, 91(2):207--210, 1991.

\bibitem[Nis96]{N96}
N.~Nisan.
\newblock {Extracting randomness: How and why: A survey}.
\newblock {\em Proceedings of the 11th Annual IEEE Conference on Computational
  Complexity}, pages 44--58, 1996.

\bibitem[Pin73]{P73}
M.~Pinsker.
\newblock On the complexity of a concentrator.
\newblock In {\em 7th Internat. Teletraffic Confer.}, pages 318/1--318/4, 1973.

\bibitem[PR97]{PR97}
S.~Popescu and D.~Rohrlich.
\newblock {Thermodynamics and the measure of entanglement}.
\newblock {\em Physical Review A}, 56(5):3319--3321, 1997.

\bibitem[Rob38]{R38}
G.~B. Robinson.
\newblock {On the Representations of the Symmetric Group}.
\newblock {\em American Journal of Mathematics}, 60(3):745--760, 1938.

\bibitem[RVW00]{RVW00}
O.~Reingold, S.~Vadhan, and A.~Wigderson.
\newblock Entropy waves, the zig-zag graph product, and new constant degree
  expanders and extractors.
\newblock In {\em FOCS}, pages 3--13, 2000.

\bibitem[Sag01]{S01}
B.~E. Sagan.
\newblock {\em {The Symmetric Group: Representations, Combinatorial Algorithms,
  and Symmetric Functions}}.
\newblock Springer, 2001.

\bibitem[Sch61]{S61}
C.~Schensted.
\newblock {Longest increasing and decreasing subsequences}.
\newblock {\em Canad. J. Math}, 13(2), 1961.

\bibitem[Ser77]{Serre77}
J.~P. Serre.
\newblock {\em Linear representations of finite groups}, volume~42 of {\em
  Graduate texts in Mathematics}.
\newblock Springer, 1977.

\bibitem[Sha02]{S02}
R.~Shaltiel.
\newblock {Recent Developments in Explicit Constructions of Extractors}.
\newblock {\em Bulletin of the EATCS}, 77:67--95, 2002.

\bibitem[SS96]{SS96}
M.~Sipser and D.~Spielman.
\newblock {Expander codes}.
\newblock {\em Information Theory, IEEE Transactions on}, 42(6):1710--1722,
  1996.

\bibitem[SV97]{SV97}
A.~Sahai and S.~Vadhan.
\newblock A complete promise problem for statistical zero-knowledge.
\newblock In {\em FOCS}, pages 448--457, 1997.

\bibitem[SV98]{SV99}
A.~Sahai and S.~Vadhan.
\newblock Manipulating statistical difference, 1998.

\bibitem[Wat02]{W02}
J.~Watrous.
\newblock Limits on the power of quantum statistical zero-knowledge.
\newblock In {\em FOCS}, pages 459--470, 2002.

\bibitem[Wat06]{W06}
J.~Watrous.
\newblock Zero-knowledge against quantum attacks.
\newblock In {\em STOC}, pages 296--305, 2006.

\end{thebibliography}

\appendix

\section{Quantum expanders from non-Abelian Cayley graphs}
\label{sec:quantum-expanders-from-non-abelian}

As we said before, our quantum expander takes two steps on a Cayley
expander (over the group \PGL) with a basis change between each of
the steps, and the basis change is a carefully chosen
transformation.

First, in Subsection \ref{sec:T}, we define and analyze taking one
step on a (Abelian or non-Abelian) Cayley graph. Then, in Subsection
\ref{sec:Abelian-expander} we analyze the Abelian case. We do not
use the results of Subsection \ref{sec:Abelian-expander} for
analyzing \PGL, but never the less we recommend reading this section
because many of its techniques are later on generalized to the
non-Abelian case. Then, we study a general template for constructing
quantum expanders over non-Abelian groups with a certain property
(Subsections \ref{sec:Template}, \ref{sec:Property}). Finally, we
show that \PGL{} has this required property (Subsection
\ref{sec:PGL}).

We begin with some representation theory background.

\subsection{Representation Theory Background}
\label{sec:rep-background}


We survey some basic elements of representation theory. For complete
accounts, consult the books of Serre \cite{Serre77} or Fulton and
Harris \cite{FH91}. The exposition below heavily uses the one given
in \cite{SRT02}.

A \emph{representation} $\rho$ of a finite group $G$ is a homomorphism
$\rho:G \to \GL(V)$, where $V$ is a (finite-dimensional) vector
space over $\C$ and $\GL(V)$ denotes the group of invertible linear
operators on $V$.  Fixing a basis for $V$, each $\rho(g)$ may be
realized as a $d \times d$ matrix over $\C$, where $d$ is the
dimension of $V$. As $\rho$ is a homomorphism, for any $g,h \in G$,
$\rho(gh)=\rho(g) \rho(h)$ (this second product being matrix
multiplication).  The \emph{dimension} $d_\rho$ of the
representation $\rho$ is $d$, the dimension of $V$.

We say that two representations $\rho_1:G \to \GL(V)$ and $\rho_2: G
\to \GL(W)$ of a group $G$ are \emph{isomorphic} when there is a
linear isomorphism of the two vector spaces $\phi:V \to W$ so that
for all $g \in G$, $\phi \rho_1(g) = \rho_2(g) \phi$. In this case,
we write $\rho_1 \cong \rho_2$. Up to isomorphism, a finite group
has a finite number of irreducible representations; we let $\widehat{G}$
denote this collection (of representations).

We say that a subspace $W \subseteq V$ is an \emph{invariant subspace}
of a representation $\rho: G \to \GL(V)$ if $\rho(g) W \subseteq W$
for all $g \in G$. The zero subspace and the subspace $V$ are always
invariant.  If no nonzero proper subspaces are invariant, the
representation is said to be \emph{irreducible}.

If $\rho: G \to \GL(V)$ is a representation, $V=V_1 \oplus V_2$ and
each $V_i$ is an invariant sub-space of $\rho$, then $\rho(g)$
defines two linear representations $\rho_i:G \to \GL(V_i)$ such that
$\rho(g)=\rho_1(g)+\rho_2(g)$. We then write $\rho=\rho_1 \oplus
\rho_2$. Any representation $\rho$ can be written $\rho = \rho_1
\oplus \rho_2 \oplus \ldots \oplus \rho_k$, where each $\rho_i$ is
irreducible. In particular, there is a basis in which every matrix
$\rho(g)$ is block diagonal, the $i$th block corresponding to the
$i$th representation in the decomposition. While this decomposition
is not, in general, unique, the \emph{number} of times a given
irreducible representation appears in this decomposition (up to
isomorphism) depends only on the original representation $\rho$.

A representation $\rho$ of a group $G$ is also automatically a
representation of any subgroup $H$. We refer to this
\emph{restricted} representation on $H$ as $\res_H \rho$.  Note that
even representations that are irreducible over $G$ may be reducible
when restricted to $H$.

The \emph{group algebra}
$\CG$ of a group $G$ is a vector space of dimension $|G|$ over $\mathbb{C}$,
with an orthonormal basis $\set{\ket{g} ~|~ g \in G}$ and
multiplication $\sum a_g \ket{g} \cdot \sum b_{g'} \ket{g'} =
\sum_{g,g'} a_g b_{g'} \ket{g \cdot g'}$. The group algebra is
isomorphic to the set $\set{f: G \to \mathbb{C}}$ with the
isomorphism being $f \to \sum_g f(g) \ket{g}$. The inner product in
$\CG$ translates to the familiar inner product $\la f, h \ra =
\sum_g \ol{f(g)} h(g)$. The \emph{regular representation} $\reg : G \to
\GL(\CG)$ is defined by $\reg(s): \ket{g} \mapsto \ket{sg}$, for any $g \in
G$.  Notice that $\reg(s)$ is a permutation matrix for any $s \in G$.

An interesting fact about the regular representation is that it
contains every irreducible representation of $G$.  In particular, if
$\rho_1,\ldots ,\rho_k$ are the irreducible representations of $G$
with dimensions $d_{\rho_1},\ldots,d_{\rho_k}$, then

$$
\reg=d_{\rho_1} \rho_1 \oplus \cdots \oplus d_{\rho_k} \rho_k,
$$
so that the regular representation contains each irreducible
representation $\rho$ exactly $d_\rho$ times.


The \emph{Fourier transform} over $G$ is a unitary transformation
$F$ mapping the standard basis $\set{\ket{g} : g \in G}$ to the
basis of the invariant subspaces of $\reg$. That is, for any $g \in
G$, the matrix $F \reg(g) F^\dagger$ is a block-diagonal matrix,
where each block corresponds to $\rho(g)$ for some irreducible
representation $\rho$ of $G$. The Fourier transform is unique, up to
a permutation of the blocks and up to a choice of basis for $\rho$
for each irreducible $\rho$.

Let $\widehat{G}$ denote the set of all inequivalent irreducible
representations of $G$. For a representation $\rho$ let $d_{\rho}$
denote the dimension of $\rho$. We define a transform $F$ by

\begin{eqnarray*}
F \ket{g} &=& \sum_{\rho \in \widehat{G}} \sum_{1 \le i,j \le
d_\rho} \sqrt{\frac{d_\rho}{|G|}} \rho_{i,j}(g) \ket{\rho,i,j}.
\end{eqnarray*}

This transformation is unique up to a choice of a unitary map
between $\Span \set{\ket{\rho,i,j} : \rho \in \widehat{G},~ 1 \le
i,j \le d_\rho}$ and $\Span \set{\ket{g} : g \in G}$.

The following analysis shows that $F$ is indeed a Fourier transform,
in the sense that it block diagonalizes the regular representations
(where each $\set{\ketbra{\rho,i,j}{\rho,i',j} : 1 \le i,i' \le
d_\rho}$ corresponds to a block).

\begin{eqnarray*} F\reg(g) F^\dagger &=& \sum_{x \in G} \sum_{\rho, \rho' \in \widehat{G}}
   \sum_{1 \le i,j \le d_\rho;~ 1 \le i',j' \le d_{\rho'}} \frac{\sqrt{d_\rho d_{\rho'}}}{|G|} \rho_{i,j}(gx) \overline{\rho'_{i',j'}(x)}
   \ketbra{\rho,i,j}{\rho',i',j'} \\
   & = & \sum_{x \in G} \sum_{\rho, \rho' \in \widehat{G}}
   \sum_{1 \le i,j \le d_\rho;~ 1 \le i',j' \le d_{\rho'}} \frac{\sqrt{d_\rho d_{\rho'}}}{|G|} \sum_{1\le k \le d_\rho} \rho_{i,k}(g)\rho_{k,j}(x) \overline{\rho'_{i',j'}(x)}
   \ketbra{\rho,i,j}{\rho',i',j'} \\
   & = & \sum_{\rho, \rho' \in \widehat{G}}
   \sum_{1 \le i,j \le d_\rho;~ 1 \le i',j' \le d_{\rho'}} \sum_{1\le k \le d_\rho} \rho_{i,k}(g)
   \left(\frac{\sqrt{d_\rho d_{\rho'}}}{|G|} \sum_{x \in G} \rho_{k,j}(x) \overline{\rho'_{i',j'}(x)}
   \right)
   \ketbra{\rho,i,j}{\rho',i',j'} \\
   & = & \sum_{\rho, \rho' \in \widehat{G}}
   \sum_{1 \le i,j \le d_\rho;~ 1 \le i',j' \le d_{\rho'}} \sum_{1\le k \le d_\rho} \rho_{i,k}(g)
   \delta_{\rho,\rho'} \delta_{k,i'} \delta_{j,j'}
   \ketbra{\rho,i,j}{\rho',i',j'} \\
   & = & \sum_{\rho \in \widehat{G}}
   \sum_{1 \le i,i',j \le d_\rho} \rho_{i,i'}(g)
   \ketbra{\rho,i,j}{\rho,i',j}
\end{eqnarray*}

In the above analysis we have used the beautiful second
orthogonality relation: $\frac{\sqrt{d_\rho d_{\rho'}}}{|G|} \sum_{x
\in G} \rho_{i,j}(x) \overline{\rho'_{i',j'}(x)} = \delta_{\rho,
\rho'} \delta_{i, i'} \delta_{j, j'}$.

\subsection{A single step on a Cayley graph} \label{sec:T}

We now fix an arbitrary (Abelian or non-Abelian) group $G$ of order
$N$, and a subset $\Gamma$ of group elements closed under inverse. The
\emph{Cayley graph} associated with $\Gamma$, $C(G,\Gamma)$, is a graph over
$N$ vertices, with an edge between $(g_1,g_2)$ iff $g_1=g_2 \gamma$ for
some $\gamma \in \Gamma$. $C(G,\Gamma)$ is a regular directed graph of degree
$|\Gamma|$. Rather then thinking of
the Cayley graph as a graph, we prefer to think of it as a linear
operator over $\CG$. We associate the graph with the operator that
is its normalized adjacency matrix $M$ (the normalization is such
that the operator norm is $1$). This operator is thus
$M=\frac{1}{|\Gamma|} \sum_{\gamma\in \Gamma} \ketbra{x\gamma}{x}$ \footnote{In our
definition the generators act from the right. Sometimes the Cayley
graph is defined with left action, i.e., $g_1$ is connected to $g_2$
iff $g_1=\gamma g_2$. However, note that if we define the invertible
linear transformation $P$ that maps the basis vector $\ket{g}$ to
the basis vector $\ket{g^{-1}}$, then $PMP^{-1}=PMP$ maps $x$ to ${1
\over |\Gamma|} \sum_\gamma \ket{(x^{-1}\gamma)^{-1}}={1 \over |\Gamma|} \sum_\gamma
\ket{\gamma^{-1}x}={1 \over |\Gamma|} \sum_\gamma \ket{\gamma x}$ and so the right action
is $M$ and the left action is $PMP^{-1}$, and therefore they are
similar and in particular have the same spectrum.}.

Notice that $M = C(G,\Gamma)$ is a symmetric operator, and therefore
diagonalizes with real eigenvalues. We denote by
$\lambda_1\ge\ldots\ge\lambda_N$ the eigenvalues of $M$ with
orthonormal eigenvectors $v_1,\ldots,v_N$ (i.e., $\norm{v_i}_2=1$).
As $M$ is regular, we have $\lambda_1=1$ and $\lbar = \max_{i>1}
|\lambda_i| \le 1$.

We now define our basic superoperator $T: L(\CG) \to L(\CG)$. The
superoperator has a register $R$ of dimension $|\Gamma|$ that is
initialized at $\ket{\ol{0}}$. It does the following:

\begin{itemize}
\item
It first applies Hadamard on register $R$ (getting into the density
matrix ${1 \over |\Gamma|} \rho \tensor \sum_{\gamma,\gamma'} \ketbra{\gamma}{\gamma'}$).

\item
Then, it applies the unitary transformation $Z: \ket{g,\gamma} \to
\ket{g\gamma,\gamma}$. This transformation is a permutation over the standard
basis, and hence unitary. It is also classically easy to compute in
both directions, and therefore has an efficient quantum circuit.

\item
Finally, it measures register $R$.
\end{itemize}

Thus we have: $T(\rho) =  \Tr_{R} [~Z (I \tensor H) (\rho \tensor
\ketbra{\ol{0}}{\ol{0}}) (I \tensor H) Z^\dagger ~]$.

We begin by identifying the eigenvectors and eigenvalues of $T$. We
may think of an eigenvector $v_i \in \C^N$ as an element of $\CG$,
$\ket{v_i}=\sum_g v_i(g) \ket{g}$. We also define the linear
transformation $R: \CG \to L(\CG)$ by $R \ket{g} = \ketbra{g}{g}$.
With this notation we define:

\begin{eqnarray*}
\mu_{i,g} & = & \reg(g) (R \ket{v_i})  ~=~ \sum_{x \in G} v_i(x)
\ketbra{gx}{x}
\end{eqnarray*}

\begin{lemm}
\label{lem:eigenvectors-of-T}The vectors $\set{\mu_{i,g} ~|~
i=1,\ldots,N, g \in G}$ form an orthonormal basis of $L(\CG)$, and
$\mu_{i,g}$ is an eigenvector of $T$ with eigenvalue
$\lambda_{i,g}=\lambda_i$.
\end{lemm}

\begin{proof}

We first notice that $T(\ketbra{g_1}{g_2}) = \Tr_\Gamma [ {1 \over |\Gamma|}
\sum_{\gamma_1,\gamma_2} U \ketbra{g_1,\gamma_1}{g_2,\gamma_2} U^\dagger]
= {1 \over |\Gamma|} \sum_{\gamma} \ketbra{g_1 \gamma}{g_2 \gamma}$. \footnote{We remark
that if we think of $T$ as an operator over $\CGG$ (identifying
$\ketbra{x}{y}$ with $\ket{x,y}$) then $T$ itself is a Cayley graph
with the set of operators being $\set{(\gamma,\gamma)~|~ \gamma \in \Gamma}$.
Furthermore, if we look at $W=\set{(g,g) ~|~ g \in G}$ then $W$ is a
subgroup of $G \times G$ and $W$ is invariant under $T$. In general,
for every $(g_1,g_2) \in G \times G$, the left coset
$(g_1,g_2)W=\set{(g_1g,g_2g) ~|~ g \in G}$ is invariant under $T$.}
Now,

\begin{eqnarray*}
T(\mu_{i,g}) & = & T(\sum_{x} v_i(x) \ketbra{gx}{x}) ~=~  \sum_{x}
v_i(x) T(\ketbra{gx}{x})\\
&=& {1 \over |\Gamma|} \sum_{x,\gamma} v_i(x) \ketbra{g x\gamma}{x\gamma} ~=~ {1 \over
|\Gamma|} \sum_{x,\gamma} v_i(x) \reg(g) \ketbra{x\gamma}{x\gamma}\\
& = & \reg(g) {1 \over |\Gamma|} \sum_{x,\gamma} v_i(x) \ketbra{x\gamma}{x\gamma} ~=~
\reg(g) R( \sum_{x} v_i(x) {1 \over |\Gamma|} \sum_{\gamma} \ket{x\gamma}) \\
& = & \reg(g) R( \sum_{x} v_i(x) M\ket{x}) \\
& = & \reg(g) R( M(\sum_{x} v_i(x) \ket{x})) ~=~ \reg(g) R( M \ket{v_i})  \\
& = & \reg(g) \cdot R(\lambda_i \ket{v_i}) ~=~ \lambda_i \reg(g)
R(\ket{v_i}) ~=~ \lambda_i \mu_{i,g}.
\end{eqnarray*}

To see orthonormality notice that for $g_1 \neq g_2$, $\Tr(\mu_{i,g}
\mu_{i',g'}^\dagger)=0$ simply because for all $(k,\ell)$ for at
least one of the matrices the $(k,\ell)$ entry is zero. If
$g_1=g_2=g$ then $\Tr(\mu_{i,g} \mu_{i',g'}^\dagger)=\la v_{i'} |
v_{i} \ra = \delta_{i,i'}$. As the number of vectors
$\set{\mu_{i,g}}$ is $N^2$ they form an orthonormal basis for
$L(\CG)$.
\end{proof}

Given $v \in \CG$ we can decompose it and express it as
$v=v^{||}+v^{\perb}$ where $v^{||} \in \Span \set{\ket{v_1}}$ and
$v^{\perb} \in \Span \set{ \ket{v_2},\ldots,\ket{v_N}}$. In analogy,
for $A \in L(\CG)$ we can decompose it to $A=A^{||}+A^{\perb}$ where
$A^{||} \in \mu^{||} = \Span \set{\mu_{1,g} ~|~ g \in G}$ and
$A^{\perb} \in \mu^{\perb} = \Span \set{\mu_{i,g} ~|~ i \neq 1, g
\in G}$. Notice that $T$ has eigenvalue $\lambda_i$ on $\mu_{i,g}$
and so in particular has eigenvalue $1=\lambda_1$ on $\mu^{||}$.
Also, let us denote $\lbar=\max_{i \neq 1} |\lambda_i|$. We have:

\begin{claim}
\label{cl:perb} For any $A \in \mu^{\perb}$, $\norm{T(A)}_2^2 \le
\lbar^2 \norm{A}_2^2$.
\end{claim}

\begin{proof}
Express $A=\sum_{i \neq 1,g} \beta_{i,g} \mu_{i,g}$. Then $
\norm{A}_2^2 = \sum_{i \neq 1,g} |\beta_{i,g}|^2$ and $ T(A) =
\sum_{i \neq 1,g} \beta_{i,g} \lambda_i \mu_{i,g}$. In particular,
$\norm{T(A)}_2^2 = \sum_{i \neq 1,g} |\beta_{i,g}|^2 |\lambda_i|^2
\le \lbar^2 \norm{A}_2^2$.
\end{proof}

\subsection{The Abelian Expander} \label{sec:Abelian-expander}

In this section we describe a quantum expander based on a Cayley
graph of an Abelian group, $G$. When $G$ is Abelian, all the
irreducible representations are of dimension $1$ and these are the
group characters \footnote{A character is a homomorphism from $G$ to
$\mathbb{C}$, .i.e., a function $\chi:G \to \mathbb{C}$ such that
$\chi(g_1 g_2)=\chi(g_1) \chi(g_2)$.}. There are exactly $N$
different characters, and we can associate each $g \in G$ with a
character $\chi_g$ such that $\chi_{g}(x)=\chi_x(g)$.
%
We associate each character $\chi$ with the norm one vector
$\ket{\chi_g} = {1 \over \sqrt{N}} \sum_x \chi_g(x) \ket{x}$ in
$\CG$.
The eigenvectors of the Cayley graph are exactly the set of
characters $\ket{v_g} = \ket{\chi_g}$.

We now describe the quantum expander. We let $U$ be the Fourier
transform over $G$, i.e., the unitary transformation mapping
$\ket{g}$ to $\ket{\chi_g}$. Our expander is the superoperator
$$E(\rho)= T( U T(\rho) U^\dagger).$$


We claim:

\begin{claim}
\label{cl:V} $U \mu_{g,i} U^\dagger =  \chi_i(g^{-1}) \cdot
\mu_{i,g^{-1}}$.
\end{claim}

\begin{proof}
\begin{eqnarray*}
U \mu_{g,i} U^\dagger & = & U \reg(i) R \ket{\chi_g} U^\dagger ~=~
{1 \over \sqrt{N}} \sum_x
\chi_g(x) U\ketbra{ix}{x}U^\dagger \\
& = & {1 \over \sqrt{N}} \sum_x \chi_g(x)
\ketbra{\chi_{ix}}{\chi_{x}}
\\
& = & {1 \over N\sqrt{N}} \sum_{x,y,y'} \chi_g(x) \chi_{ix}(y)
\ol{\chi_{x}(y')} \ketbra{y}{y'} \\
& = & {1 \over \sqrt{N}} \sum_{y,y'} \chi_i(y) ~~[~{1 \over N}
\sum_x \chi_x(gyy'^{-1})~]~ \ketbra{y}{y'} \\
& = & {1 \over \sqrt{N}} \sum_{y'} \chi_i(g^{-1}y') \ketbra{g^{-1}y'}{y'} \\
 & =& \chi_i(g^{-1}) \cdot \reg(g^{-1}) R \ket{\chi_i} ~=~ \chi_i(g^{-1}) \cdot \mu_{i,g^{-1}}
\end{eqnarray*}
\end{proof}

We claim:

\begin{lemm}
$E$ is a $(|\Gamma|^2, \lbar)$ quantum expander.
\end{lemm}

\begin{proof}
The regularity is clear from the way the superoperator $E$ is
defined. We turn to the spectral gap. It is easy to check that
$E(\nI)=\nI$. Furthermore, fix any $x \in L(\CG)$ that is
perpendicular to $\nI$. Write $x=x^{||}+x^{\perb}$ where $x^{||} \in
W=\Span \set{\mu_{1,g} ~|~ 1 \neq g \in G}$ and $x^{\perb} \in
\mu^{\perb}$. Given Claim \ref{cl:V} one can verify that $E(x^{||})
\perb E(x^{\perb})$. In particular

\begin{eqnarray*}
||E(x)||_2^2 & = & ||E(x^{||})||_2^2 + ||E(x^{\perb})||_2^2 \\
& \le &  ||T(U x^{||} U^\dagger)||_2^2 + ||T(x^\perb)||_2^2 \\
& \le & \lbar^2 ||x^{||}||_2^2+\lbar^2 ||x^\perb||_2^2 ~=~ \lbar^2
||x||_2^2.
\end{eqnarray*}

The first inequality is due to the fact that $T$ has eigenvalue $1$
on $x^{||}$ and both $T$ and $U$ have operator norm at most $1$. The
second inequality is by Claims \ref{cl:V} and \ref{cl:perb}.
\end{proof}

\subsection{Template for a quantum expander over a general group}
\label{sec:Template}

In this subsection we show how to construct a quantum expander over
any group $G$ that possess some general property. We later show that
the $\PGL$ group possesses this property.

Similar to the Abelian case, the expander will be of the form
$$E(\rho)= T( U T(\rho) U^\dagger),$$ where $U$ will be the Fourier
transform over $G$. Unlike the Abelian case, in the non-Abelian case
$G$ has many representations of dimension greater than $1$. Thus, a
significant part of describing $U$ will be to describe the basis for
each one of the $\reg$-invariant subspaces. The property that we
need from the unitary transformation $U$ is:

\begin{deff}
We say $U$ is a \emph{good basis change} if for any $g_1 \ne e$
(where $e$ denotes the identity element) it holds that

\begin{eqnarray}
\label{eqn:propertyU} \Tr(U \reg(g_1) U^\dagger \reg(g_2)) &=& 0.
\end{eqnarray}
\end{deff}

The intuition behind this choice is as follows. As before, let $W =
\Span \set{\reg(g) : g \ne e \in G}$ be the set of eigenvectors of
$T$ with eigenvalue $1$ (besides the identity). Since each of these
eigenvectors was not shrunk by $T$ in the first step, it is
necessary to move them into a perpendicular subspace, such that the
second step will shrink them. If $U$ is a good basis change this
indeed happens as captured in:

\begin{claim}
\label{cl:sigma||} If $\rho \in W$ and $U$ is a good basis change
then $U \rho U^\dagger \perp \mu^{||}$ (i.e. $U \rho U^\dagger \in \mu^{\perp}$).
\end{claim}

\begin{proof}
$\set{\reg(g) : g \in G}$ is an orthonormal basis for $\mu^{||}$.
$\set{\reg(g) : g \ne e \in G}$ is an orthonormal basis for $W$.
Therefore, it is enough to verify that $\Tr(U \reg(g_1) U^\dagger
\reg(g_2)^\dagger) = 0$ for any $g_1 \ne e$ and for any $g_2$. Since
$\reg(g_2)^\dagger = \reg(g_2^{-1})$, this follows directly from
Property (\ref{eqn:propertyU}).
\end{proof}

We claim:

\begin{lemm}\label{lem:Expander template}
If $U$ is a good basis change then $E$ is a $(|\Gamma|^2, \lbar)$ quantum
expander.
\end{lemm}

\begin{proof}
The regularity is clear from the way the superoperator $E$ is
defined. We turn to the spectral gap. It is easy to check that
$E(\nI)=\nI$. Furthermore, fix any $x \in L(\CG)$ that is
perpendicular to $\nI$. Write $x=x^{||}+x^{\perb}$ where $x^{||} \in
W=\Span \set{\mu_{1,g} ~|~ e \neq g \in G}$ and $x^{\perb} \in
\mu^{\perb}$. Now it is not true any more that $E(x^{||}) \perb
E(x^{\perb})$. However, $E(x)= T(\sigma^{||}+\sigma^{\perb})$, where
$\sigma^{||}=UT(x^{||})U^\dagger$ and
$\sigma^{\perb}=UT(x^\perb)U^\dagger$. We know a few things. First,
by Claim \ref{cl:sigma||}, $\sigma^{||} \perb \mu^{||}$. Also,
$T(x^{||}) \perb T(x^\perb)$, and therefore $\sigma^{||} \perb
\sigma^\perb$. Finally, by Lemma \ref{lem:eigenvectors-of-T} we know
$T$ is normal. We soon prove:

\begin{lemm}
\label{lem:col-sum} Let $T$ be a normal linear operator with
eigen-spaces $V_1,\ldots,V_n$ and corresponding eigenvalues
$\lambda_1, \ldots ,\lambda_n$ in descending \emph{absolute} value.
Suppose $u$ and $w$ are vectors such that $u \in \Span
\set{V_2,\ldots,V_n}$ and $w \perp u$ ($w$ does not necessarily
belong to $V_1$). Then

\begin{eqnarray*}
||(T(u + w))||_2^2 \le |\lambda_2|^2 ||u||_2^2 + |\lambda_1|^2
||w||_2^2.
\end{eqnarray*}
\end{lemm}

Using the lemma we see that:

\begin{eqnarray*}
||E(x)||_2^2 & = & ||T(\sigma^{||}+\sigma^{\perb})||_2^2 \\
& \le & \lbar^2 ||\sigma^{||}||_2^2 + ||\sigma^\perb||_2^2 \\
&=& \lbar^2 ||UT(x^{||})U^\dagger||_2^2
+||UT(x^\perb)U^\dagger||_2^2 \\
&=& \lbar^2 ||T(x^{||})||_2^2+ ||T(x^\perb)||_2^2 \\
 & \le
& \lbar^2 ||x^{||}||_2^2+ \lbar^2 ||x^\perb||_2^2 ~=~
 \lbar^2 ||x||_2^2.
\end{eqnarray*}
\end{proof}

We are left with the proof of Lemma \ref{lem:col-sum}:

\begin{proof}(Of Lemma \ref{lem:col-sum})
Let $\set{v_j}$ be an eigenvector basis for $T$ with eigenvalues
$\delta_j$ (from the set $\set{\lambda_1,\ldots,\lambda_n}$).
Writing $u = \sum_{j} \alpha_j v_j$ and $w = \beta v +\sum_{j}
\beta_j v_j$ with $v_j \in \Span \set{V_2,\ldots,V_n}$ and $v \in
V_1$, we get:

\begin{eqnarray*}
||T(u + v)||_2^2 & = & ||\lambda_1 \beta v + \sum_{j}
\delta_j(\alpha_j + \beta_j) v_j||_2^2 \\
& = & |\lambda_1|^2 |\beta|^2 + \sum_{j} |\delta_j|^2|\alpha_j + \beta_j|^2 \\
& \le & |\lambda_1|^2 |\beta|^2 + |\lambda_2|^2 \sum_{j} |\alpha_j + \beta_j|^2 \\
& = & |\lambda_1|^2 |\beta|^2 + |\lambda_2|^2 (\sum_{j} |\alpha_j|^2 + \sum_{j} |\beta_j|^2 + \sum_{j} (\alpha_j^* \beta_j + \alpha_j \beta_j^*)) \\
& = & |\lambda_1|^2 |\beta|^2 + |\lambda_2|^2 (\sum_{j} |\alpha_j|^2 + \sum_{j} |\beta_j|^2) \\
& \le & |\lambda_1|^2 (|\beta|^2+\sum_{j} |\beta_j|^2) +
|\lambda_2|^2 \sum_{j} |\alpha_j|^2 = |\lambda_2|^2 ||u||_2^2 +
|\lambda_1|^2 ||w||_2^2.
\end{eqnarray*}

where in the calculation we used the fact that $\sum_{j} \alpha_j^*
\beta_j = \la u | v \ra =0$ because of the orthogonality of $u$ and
$w$.
\end{proof}

\subsection{A sufficient condition that guarantees a good basis change} \label{sec:Property}


\begin{deff}
Let $f$ be a bijection from $\set{(\rho,i,j)~|~ \rho \in \wh{G}, 1
\le i,j \le d_\rho}$ to $G$.  We say that $f$ is \emph{product} if
for every $\rho \in \wh{G}$:

\begin{eqnarray} \label{eqn:product} f(\rho,i,j) &=& f_1(i) \cdot
f_2(j)
\end{eqnarray}
for some functions $f_1,f_2: [d_\rho] \times [d_\rho] \to G$ ($f_1$
and $f_2$ may depend on $\rho$).
\end{deff}

We first give two examples.

\begin{example}\label{example:product:abelian} (Abelian groups). All irreducible representations are of dimension one, so
just define $f_1(i)=e$ and $f_2(j)=f(\rho,1,1)$.
\end{example}

\begin{example}\label{example:product:dihedral}(The Dihedral group)
The Dihedral group $D_m$ is the group of rotations and reflections
of a regular polygon with $m$ sides. Its generators are $r$, the
rotation element, and $s$, the reflection element. This group has
$2m$ elements and the defining relations are $s^2 = 1$ and $srs =
r^{-1}$. We shall argue this group has a product mapping for odd $m$
(although it is true for even $m$ as well). The Dihedral group has
$\frac{m-1}{2}$ representations $\rho_\ell$ of dimension two and two
representations of dimension one $\tau_1,\tau_2$.

Our product mapping $f(\rho,i,j)$ is:

\begin{eqnarray}
\label{eqn:dihedral} f(\rho,i,j) & = & \left\{
\begin{array}{ll}
1 & \mbox{If $\rho=\tau_1,i=j=1$} \\
s & \mbox{If $\rho=\tau_2,i=j=1$} \\
r^{2(\ell-1)+i}s^j & \mbox{If $\rho=\rho_\ell$}
\end{array}
\right.
\end{eqnarray}

The product structure is clear from Equation (\ref{eqn:dihedral}).
%
\end{example}

Our claim is that any group that has a product mapping can be used
to construct quantum expanders. The parameters of the expander
depend on the parameters of the classical Cayley graph given by the
group. Optimally, we will want a group that has:

\begin{itemize}
\item A constant degree Cayley expander.

\item A product mapping.

\item An efficient quantum Fourier transform.
\end{itemize}

Abelian groups have the last two. In the next section we will show
that \PGL{} has the first two (it is an open problem to find an
efficient implementation of the quantum Fourier transform over
\PGL{}).

\begin{lemm}\label{lem:product implies good basis}
Let $G$ be a group that has a product mapping $f$, and let $F$ be
the Fourier transform over $G$, $F \ket{g} = \sum_{\rho \in
\widehat{G}} \sum_{1 \le i,j \le d_\rho} \sqrt{\frac{d_\rho}{|G|}}
\rho_{i,j}(g) \ket{\rho,i,j}$. Define the unitary mapping

\begin{eqnarray*}
S &:& \ket{\rho,i,j} \mapsto \omega_{d_\rho}^{ij} \ket{f(\rho,i,j)}
\end{eqnarray*}

where $\omega_{d_\rho} = e^{{2\pi i}/{d_\rho}}$, and set $U$ to be
the unitary transformation $U=SF$. Then $U$ has property
$(\ref{eqn:propertyU})$ and is a good basis change.
%
\end{lemm}


\begin{proof}

\begin{eqnarray*}
\Tr\left(U \reg(g_1) U^\dagger \reg(g_2)\right) & = & \Tr\left(SF
\reg(g_1) F^\dagger S^\dagger \reg(g_2)\right) \\
& = & \Tr\left(S\sum_{\rho \in \widehat{G}} \sum_{1 \le i,i',j \le
d_\rho} \rho_{i,i'}(g_1) \ketbra{\rho,i,j}{\rho,i',j} S^\dagger
\sum_{x}
\ketbra{g_2 x}{x}\right) \\
& = & \sum_{\rho \in \widehat{G}} \sum_{1 \le i,i' \le d_\rho}
\rho_{i,i'}(g_1) \Tr\left(\sum_{j=1}^{d_\rho}
S\ketbra{\rho,i,j}{\rho,i',j}S^\dagger \sum_{x} \ketbra{g_2
x}{x}\right).
\end{eqnarray*}

Therefore, it suffices to show that for any $\rho, i, i'$ we have
$\Tr\left(\sum_{j=1}^{d_\rho} S\ketbra{\rho,i,j}{\rho,i',j}S^\dagger
\sum_{x} \ketbra{g_2 x}{x}\right) = 0$. Fix $\rho \in \wh{G}$ and
$i,i' \in \set{1,\ldots,d_\rho}$. Since $f$ is product, $f(\rho,i,j)
= f_1(i) \cdot f_2(j)$ for some $f_1,f_2: [d_\rho] \times [d_\rho]
\to G$. Denote $h_i=f_1(i)$ and $t_j=f_2(j)$. The sum we need to
calculate can be written as

\begin{eqnarray*}
\Tr\left(\sum_{j=1}^{d_\rho} S\ketbra{\rho,i,j}{\rho,i',j}S^\dagger
\sum_{x} \ketbra{g_2 x}{x}\right) &=& \sum_{j=1}^{d_\rho} \sum_{x}
\omega_{d_\rho}^{ij - i'j} \Tr\left(\ket{h_i t_j} \braket{h_{i'}
t_j}{g_2 x}\bra{x}\right) \\
& = & \sum_{j=1}^{d_\rho} \sum_{x} \omega_{d_\rho}^{ij - i'j}
\braket{x}{h_i t_j} \braket{h_{i'} t_j}{g_2 x}
\\
&=& \sum_{j=1}^{d_\rho} \omega_{d_\rho}^{(i - i')j}
\braket{g_2}{h_{i'} h_{i}^{-1}}.
\end{eqnarray*}

where the last equality is because we get a non-zero value iff
$x=h_i t_j$ and $h_{i'}t_j=g_2x$, which happens iff $h_i
t_j=g_2^{-1}h_{i'}t_j$, i.e.,
$g_2=h_{i'} h_{i}^{-1}$. However, when $g_2=h_{i'} h_{i}^{-1}$ we
get the sum $\sum_{j=1}^{d_\rho} \omega_{d_\rho}^{(i - i')j}$. This
expression itself is zero when $i \neq i'$.

We are therefore left with the case $i=i'$. In this case $g_2=h_{i'}
h_{i}^{-1}=e$. But then,

\begin{eqnarray*}
\Tr\left(U \reg(g_1) U^\dagger \reg(g_2)\right) & = & \Tr\left(U
\reg(g_1) U^\dagger \right) = \Tr\left( \reg(g_1) \right) = 0,
\end{eqnarray*}

where the last equality follows because  $g_1 \ne e$.
\end{proof}

\subsection{The construction of the \PGL{} quantum expander} \label{sec:PGL}

We now work with the group $G = \PGL$ of all $2 \times 2$ invertible
matrices over $\F_q$ modulo the group center (the set of scalar
matrices). This is one of the groups used by \cite{LPS88} to
construct Ramanujan expander graphs. Our goal is to show that $\PGL$
has a product mapping. We therefore need to show a product bijection
between $G$ and the irreducible representations of $G$. How can we
find such a bijection?

We first describe the well known irreducible representations of this
group. These are:

\begin{itemize}
\item $\frac{q-3}{2}$ representations of dimension $q+1$.
\item $\frac{q-1}{2}$ representations of dimension $q-1$.
\item 2 representations of dimension $q$.
\item 2 representations of dimension $1$.
\end{itemize}

We need a clean bijection from $G$ to the irreducible
representations of $G$. Our approach is to use a tower of subgroups,
$G_3=G > G_2= D_{2q} > G_1=Z_q > G_0=\set{e}$, with $G_2$ and $G_1$
defined as follows. $G_2$ is generated by the equivalence classes of
$\left(%
\begin{array}{cc}
  1 & 0 \\
  0 & -1 \\
\end{array}%
\right)$
and of $\left(%
\begin{array}{cc}
  1 & 1 \\
  0 & 1 \\
\end{array}%
\right).$ $G_2$ is a Dihedral subgroup of $G$ with $2q$ elements.
The first matrix is the reflection, denoted by $s$, and the second
is the rotation, denoted by $r$. This group has a cyclic subgroup
$G_1 = Z_q$ (the group generated by $r$).

\begin{figure}
  \includegraphics[width=0.8\textwidth]{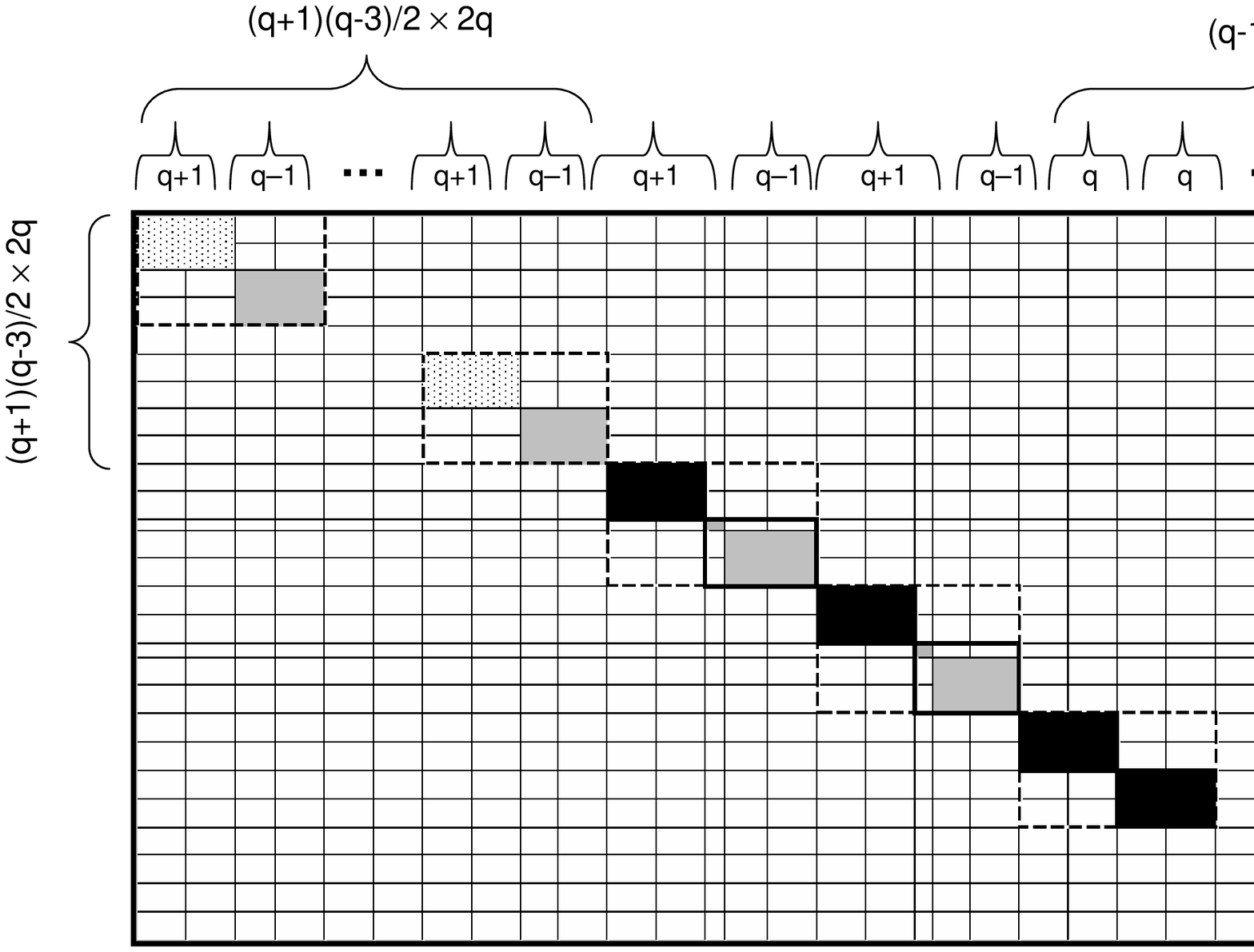}\\
  \caption{The product mapping of \PGL.}\label{picture}
\end{figure}

In Figure \ref{picture} we show the product mapping visually. The
figure shows the block-diagonal structure of the regular
representation (after applying the Fourier transform). Each
rectangle is an irreducible representation. Each color represents a
different dimension: black rectangles correspond to irreducible
representations of dimension $q$, gray rectangles correspond to
irreducible representations of dimension $q-1$ and dotted rectangles
correspond to irreducible representations of dimension $q+1$. Notice
that all rectangles fit into larger block diagonal rectangles of
dimension $2q$, marked with dashed lines. These larger rectangles
correspond to cosets of $G_2$. It is straightforward to verify that
for any $q+1$ dimensional representation (dotted rectangles in the
figure), the product condition is satisfied by $G_2$, by letting
$f_1(i)$ determine the index in the coset, and $f_2(j)$ determine
the coset representative. Similarly, for any other representation
(black and gray rectangles in the figure) the product condition is
satisfied by $G_1$.

Formally, our product mapping $f$ is defined as follows. Let $\ell =
\frac{(q-1)(q+1)}{2}$ and let $T_2 = \set{t_1, \ldots, t_\ell}$ be a
transversal for $G_2$ (its size comes from the fact that $|G| =
q(q-1)(q+1) = 2q \ell$). $T_1 = \set{t_1, s t_1, \ldots, t_\ell, s
t_\ell}$ is a transversal for $G_1$. We denote by $\rho^d_x$ the
$x$th representation of dimension $d$ (these are all non-equivalent
irreducible representations).

\begin{eqnarray*}
f(\rho^1_x,1,1) &=& s t_{x + \frac{(q-3)(q+1)}{2}} \\
f(\rho^{q-1}_x,i,j) &=& r^i s t_{(x-1)(q-1) + j} \\
f(\rho^q_x,i,j) &=& \left\{
\begin{array}{ll}
    r^{i-1}   t_{(x-1)q + j + \frac{(q-3)(q+1)}{2}} & (x-1)q + j \le q+1 \\
    r^{i-1} s t_{(x-1)q + j -q+1 + \frac{(q-3)(q+1)}{2}} & \text{otherwise} \\
\end{array}
\right. \\
f(\rho^{q+1}_x,i,j) &=& \left\{
\begin{array}{ll}
    r^{i-1} t_{(x-1)(q+1) + j} & i \le q \\
          s t_{(x-1)(q+1) + j} & i = q+1 \\
\end{array}
\right.
\end{eqnarray*}

As we see, for every $\rho \in \wh{G}$, $f$ has a product structure.

We remark that the two previous examples of product mappings
(Examples \ref{example:product:abelian} and
\ref{example:product:dihedral}) have also this subgroup structure.
In the Abelian case (Example \ref{example:product:abelian}) we use
$G < G_0=\set{e}$ and in the dihedral case $G=D_{2n} < G_1= Z_n <
G_0=\set{e}$ (or alternatively,  $G=D_{2n} < G_1= \set{e,s} <
G_0=\set{e}$).

We are now ready to prove Theorem \ref{thm:PGL-expander}.

\newtheorem*{thmb}{Theorem~\ref{thm:PGL-expander}}
\begin{thmb}\label{thm:PGL-expander}
There exists a $(D=O({1 \over \lbar^4}),\lbar)$ quantum expander.
\end{thmb}

\begin{proof}
By Lemma \ref{lem:Expander template}, Lemma \ref{lem:product implies
good basis} and the description of the product mapping above, we
know that $E$ is a $(|\Gamma|^2, \lbar)$ quantum expander. By the
\cite{LPS88} construction we know that there exists a Cayley graph
for $\PGL$ with $\lbar^2 \le {4 \over |\Gamma|}$. Plugging this Cayley
graph gives us a $(\frac{16}{\lbar^4}, \lbar)$ quantum expander.
\end{proof}

\subsection{How about an $\Sn$ Cayley construction ?}
\label{sec:Sn-construction}

There are explicit, constant degree (non-Ramanujan) Cayley expanders
over $\Sn$ \cite{K05}. Also, there is an efficient implementation of
the Fourier transform over $\Sn$ \cite{B97}. We do not know,
however, whether $\Sn$ has product mappings. In $\Sn$ the question
takes the following form. We look for bijections $f$ from pairs
$(P,T)$ of standard shapes to $\Sn$ (a shape corresponds to an
irreducible representation of $\Sn$, and its dimension is the number
standard shapes of that shape), and furthermore we want $f(P,T)$ to
equal $f_1(P) \cdot f_2(T)$ for some functions $f_1$ and $f_2$
taking values in $\Sn$ (this is the product property).

The question of finding an explicit bijection $f$ from pairs $(P,T)$
of standard shapes to $\Sn$ is a basic question in the study of the
representation theory of $\Sn$. The canonical algorithm doing so is
the "Robinson-Schensted" algorithm \cite{R38,S61} that was
extensively studied later on (see \cite{S01}, and especially Chapter
3 that is almost dedicated to this algorithm). The R-S mapping is
\emph{not} product. However, a manual check revealed that $S_n$ has
a product mapping for $n \le 6$. We think it is a natural question
whether product mappings for $\Sn$ exist, and if so it is natural to
look for an explicit description of the mapping (preferably by an
algorithm).

\section{The complexity of estimating entropy}
\label{sec:estimating-entropy}

In this section we show that the $\QED$ problem (as defined in the
introduction) is $\QSZK$-complete, following the outline in Section
\ref{sec:main:estimating-entropy}. We prove that $\QEA \le
\ol{\QSD}$ in Section \ref{sec:QEA-QSD} and  that $\QEA \le \QSD$
implies $\QED \le \QSD$ in Section \ref{sec:closure}. This completes
the proof that $\QED \le \QSD$. We prove that $\QSD \le \QED$ in
Section \ref{sec:QSD-QED}.

Before we begin we need a few things. First we need a well known
fact about the trace-norm. In the classical world
$\text{SD}(P,Q)=\max_S P(S)-Q(S)$, i.e., it describes the maximal
probability with which one can distinguish the two distributions.
The trace distance achieves the same for density matrices, as is
captured in:

\begin{fact}(e.g., \cite{NC00})\label{fact:trn}
Let $\rho_0$ and $\rho_1$ be two density matrices. Then there exists
a measurement $\mathcal{O}$ with outcome $0$ or $1$ such that making
the measurement on $\rho_b$ yields the bit $b$ with probability
$\half + \frac{\trn{\rho_0 - \rho_1}}{2}$. Furthermore, no
measurement can distinguish the two density matrices better.
\end{fact}

As with classical distributions, the distance between density
matrices can only decrease with computation, i.e.,

\begin{fact} (\cite{NC00}) \label{fact:quantum-operator-decrease-distance}
Let $\rho_0$ and $\rho_1$ be two density matrices. Then for any
quantum operation $\mathcal{E}$ it holds that
$\trn{\mathcal{E}(\rho_0) - \mathcal{E}(\rho_1)} \le \trn{\rho_0 -
\rho_1}$.
\end{fact}

We also need the polarization lemma \cite{W02} (that is based on the
work of \cite{SV97}), which is used throughout the section.

\begin{thm}
\label{thm:polarization} \textbf{(Polarization lemma, Theorem 5 at
\cite{W02})} Let $\alpha$ and $\beta$ satisfy $0 \le \alpha <
\beta^2 \le 1$. Then there is a deterministic polynomial-time
procedure that, on input $(Q_0, Q_1, 1^n)$ where $Q_0$ and $Q_1$ are
quantum circuits, outputs descriptions of quantum circuits $(R_0,
R_1)$ (each having size polynomial in $n$ and in the size of $Q_0$
and $Q_1$) such that

\begin{eqnarray*}
\trnorm{\ket{Q_0} - \ket{Q_1}} \le \alpha & \Rightarrow &
\trnorm{\ket{R_0} - \ket{R_1}} \le 2^{-n}, \\
\trnorm{\ket{Q_0} - \ket{Q_1 }} \ge \beta & \Rightarrow &
\trnorm{\ket{R_0} - \ket{R_1}} \ge 1-2^{-n}.
\end{eqnarray*}
\end{thm}

\subsection{$\QEA \le \ol{\QSD}$}
\label{sec:QEA-QSD}

In Section \ref{sec:main:estimating-entropy} we gave an intuitive
explanation of what follows. We first prove the quantum version of
the flattening lemma (Lemma \ref{lem:flattening}), then describe the
reduction, and argue that if the input source had much entropy we
are close to uniform, whereas if the input source had few entropy,
then applying the extractor does not add much entropy, and that this
implies that the output state is far away from the completely mixed
state (Lemma \ref{lem:few-entropy-is-far-from-uniform}).

\begin{deff}
Let $\rho$ be a density matrix, $\lambda$ an eigenvalue of $\rho$
and $\Delta$ a positive number. We say that $\lambda$ is
\emph{$\Delta$-typical} if $2^{-S(\rho) -\Delta} \le \lambda \le
2^{-S(\rho) + \Delta}$. We say $\rho$ is \emph{$\Delta$-flat} if for
every $t>0$, with probability $\ge 1-2^{-t^2 + 1}$, a measurement of
$\rho$ in its eigenvector basis results with an eigenvector whose
eigenvalue is $t \Delta$-typical .
\end{deff}

\begin{lemm} \label{lem:flattening}
Let $\rho$ be a density matrix and $k$ a positive integer. Suppose
that every non-zero eigenvalue of $\rho$ is at least $2^{-m}$. Then
$\tensor ^k \rho$ is $\Delta$-flat for $\Delta = \sqrt{k}m$.
\end{lemm}
\begin{proof}
Let $\set{\lambda_1,\ldots,\lambda_n}$ denote the set of eigenvalues
of $\rho$. This implies the eigenvalues of $\tensor ^k \rho$ are
$\set{\lambda_{i_1,\ldots,i_k} ~:~ \lambda_{i_1,\ldots,i_k} =
\lambda_{i_1} \cdot \ldots \cdot \lambda_{i_k} }$. The entropy of
$\tensor ^k \rho$ is $S(\tensor ^k \rho) = k \cdot S(\rho)$. Let $A$
denote the set of $t \Delta$-typical eigenvalues of $\tensor ^k
\rho$. Thus $A = \set{\lambda_{i_1,\ldots,i_k} : |\sum_{j=0}^k -\log
\lambda_{i_j} - k \cdot S(\rho)| \le t \Delta}$. Let $p$ denote the
probability that a measurement of $\tensor ^k \rho$ in its
eigenvector basis results an eigenvalue which is not $t
\Delta$-typical. Then by Hoeffding inequality,
$$ p \le \sum_{x \notin A} x \le 2 \exp \left( \frac{-2 \cdot k \cdot (t\Delta/k)^2}{m^2} \right) \le 2 \exp(-2t^2) \le 2^{-t^2 +1}.$$
\end{proof}

We now define the reduction. Let $(Q,t)$ be an input to $\QEA$,
where $Q$ is a quantum circuit with $n$ input qubits and $m$ output
qubits. We first look at the circuit $Q^{\tensor q}$ (for some $q =
\poly(n)$ to be specified later). We let $E$ be a $(qt,q(m-t) +
2\logeps + \log (qm) + O(1),\epsilon)$ quantum extractor operating
on $qm$ qubits, where $\epsilon = 1 / \poly(n)$ will be fixed later.
Such an extractor exists by Lemma \ref{lem:expander-extractor}. Let
$\xi = E(\ket{Q}^{\tensor q})$ and let $\nI=2^{-qm}I$. The output of
the reduction is $(\xi,\nI)$.

To show correctness we prove:

\begin{lemm}
\begin{itemize}
\item
If $(Q,t) \in \QEA_Y$ then $\trn{\xi - \nI} \le 5 \epsilon$.
\item
If $(Q,t) \in \QEA_N$ then $\trnorm{\xi - \nI} \ge \frac{1}{qm} -
\frac{1}{2^{qm}}$. \end{itemize}
\end{lemm}

\begin{proof}

\begin{itemize}
\item Proof of the first item.

Since $Q$ traces out at most $n$ qubits, the eigenvalues of
$\ket{Q}$ are all at least $2^{-n}$, and by Lemma
\ref{lem:flattening} we see that $\ket{Q}^{\tensor q}$ is
$\Delta$-flat for $\Delta = \sqrt{q}n$. Thus, with probability at
least $1-2^{-r^2 + 1}$, a measurement of $\ket{Q}$ in its
eigenvector basis results with an eigenvector whose eigenvalue is $r
\Delta$-typical. Let $\Lambda$ denote the set of $r \Delta$-typical
eigenvalues of $\ket{Q}$, for $r = \sqrt{\logeps}$. We write
$\ket{Q}^{\tensor q}$ in its eigenvector basis $\ket{Q}^{\tensor q}
= \sum_i \lambda_i \ketbra{v_i}{v_i}$. Let $\sigma_0 =
\sum_{\lambda_i \in \Lambda} \lambda_i \ketbra{v_i}{v_i}$, and let
$\sigma_1 = \rho^{\tensor q} - \sigma_0$. Thus, $\Tr(\sigma_0) \ge
1-2^{-r^2 + 1}$. Therefore,

\begin{eqnarray*}
\trnorm{\xi - \nI} & =& \trnorm{E(\sigma_0)+E(\sigma_1)-\Tr(\sigma_0)\tilde{I}-\Tr(\sigma_1)\tilde{I}}   \\
&\le& \trnorm{E(\sigma_0) - \Tr(\sigma_0)\nI} + \trnorm{E(\sigma_1)}
+ \trnorm{\Tr(\sigma_1)\nI} \\ &\le&
\trnorm{E(\frac{1}{\Tr(\sigma_0)}\sigma_0) - \nI} + 2^{-r^2+2} .
\end{eqnarray*}

Now we use the fact that $\frac{1}{\Tr(\sigma_0)}\sigma_0$ is a
density matrix with all its eigenvalues $\le 2^{-q \cdot S(\rho) +
r\Delta} \cdot \frac{1}{\Tr(\sigma_0)} \le 2^{-q \cdot S(\rho) +
r\Delta + 1}$. Thus, $\frac{1}{\Tr(\sigma_0)}\sigma_0$ has
min-entropy at least $q \cdot S(\rho) - r\Delta - 1 \ge q \cdot
(t+1) - r\Delta - 1$ since we started with a yes instance for
$\QEA_Y$. We set the parameters such that $q \ge r\Delta + 1$, and
thus our density matrix has min-entropy at least $qt$ and by the
guarantee of our quantum extractor we get that
$\trnorm{E(\frac{1}{\Tr(\sigma_0)}\sigma_0) - \nI} \le \epsilon$.
Therefore, $\trnorm{\xi - \nI} \le \epsilon + 2^{-r^2 + 2} \le 5
\epsilon$, where the last inequality holds for $r \ge
\sqrt{\logeps}$.

\item Proof of the second item.

Suppose that $(Q,t) \in \QEA_N$. By the definition of quantum
extractors we get that
\begin{eqnarray*}
S(\xi) & \le & S(\ket{Q}^{\tensor q}) + q(m-t) + 2\logeps + \log(qm) + O(1) \\
& \le & q(t-1) + q(m-t) + 2\logeps + \log(qm) + O(1) \\
& = & qm - q + 2\logeps + \log(qm) + O(1) \le qm - 1,
\end{eqnarray*}
where the last inequality follows if we choose the parameters such
that $q > 2\logeps + \log(qm) + O(1)$.

Here we need to relate entropy to statistical distance. This is
given in Lemma \ref{lem:few-entropy-is-far-from-uniform}, which we
state and prove soon. With that we get that $\trnorm{\xi - \nI} \ge
\frac{1}{qm} - \frac{1}{2^{qm}}$ as required.
\end{itemize}
\end{proof}

The constraints we have on the parameters are $q \ge
\sqrt{\logeps}\sqrt{q}n + 1$ and $q > 2\logeps + \log(qm) + O(1)$.
To this we add $5 \epsilon < \left(\frac{1}{qm} -
\frac{1}{2^{qm}}\right)^2$. This ensures a gap which can be
amplified by Theorem \ref{thm:polarization} to any desired gap, and
completes the proof. These constraints can be easily satisfied by
choosing $q$ and $\epsilon^{-1}$ to be appropriately large
polynomials in $n$.

\subsubsection{Relating entropy to trace distance from the completely
mixed state} \label{sec:ent-dist}

Now we relate the distance of a density matrix from uniform to a
bound on its entropy. Consider the following classical random
variable $X$ over $\B^n$: with probability $\epsilon$, $X$ samples
the fixed string $0^n$ and with probability $1 - \epsilon$, $X$ is
uniformly distributed over $\B^n$. This $X$ has distance about
$\epsilon$ from uniform ($\epsilon + \frac{1-\epsilon}{2^n}-
\frac{1}{2^n}$ to be exact) and its entropy is  $S(\rho) \le
(1-\epsilon)n + H(1-\epsilon)$. We show that this is essentially the
worst possible:

\begin{lemm} \label{lem:few-entropy-is-far-from-uniform}
Let $\rho$ be a density matrix over $n$ qubits and $\epsilon>0$. If
$S(\rho) \le (1-\epsilon)n$ then $\trn{\rho - \frac{1}{2^n}I} \ge
\epsilon - \frac{1}{2^n}$.
\end{lemm}

\begin{proof}
We prove the contrapositive. Let $\rho$ be a density matrix with
$\trnorm{\rho - \frac{1}{2^n}I} < \epsilon - \frac{1}{2^n}$ and
minimal Shannon entropy. Writing $\rho$ in its eigenvector basis we
get $\rho = \sum_{i=1}^{2^n} \lambda_i \ketbra{v_i}{v_i}$. W.l.o.g
let us assume $\lambda_1$ is the largest eigenvalue of $\rho$. The
trace distance of $\rho$ from $\frac{1}{2^n}I$ is $\half \sum_i
|\lambda_i - \frac{1}{2^n}|$. For any eigenvalue $\lambda_i >
\frac{1}{2^n}$, where $i \ne 1$, we can modify the eigenvalues of
$\rho$ such that $\lambda_1 \leftarrow \lambda_1 + (\lambda_i -
\frac{1}{2^n})$ and $\lambda_i \leftarrow \frac{1}{2^n}$. Since both
$\lambda_1$ and $\lambda_i$ are $\ge \frac{1}{2^n}$, this does not
affect $\trnorm{\rho - \frac{1}{2^n}I}$. Moreover, we claim this
operation only decreases $S(\rho)$:

\begin{lemm}
\label{lem:local-change} Let $\rho = \sum_{i=1}^{2^n} \lambda_i
\ketbra{v_i}{v_i}$ be a density matrix over $n$ qubits with
eigenvalues $(\lambda_1 \ge \ldots \ge \lambda_{2^n})$. Let
$\lambda_j > \epsilon > 0$ for some $j>1$. Let $\delta_1 = \lambda_1
+ \epsilon$, $\delta_j = \lambda_j - \epsilon$ and $\delta_i =
\lambda_i$ for $i \ne 1,j$ and let $\sigma = \sum_{i=1}^{2^n}
\delta_i \ketbra{v_i}{v_i}$. Then $S(\rho) \ge S(\sigma)$.
\end{lemm}

We prove the lemma shortly. Thus, w.l.o.g. we can assume $\lambda_i
\le 2^{-n}$ for all $i>1$. Having that $\trn{\rho-\tilde{I}} =
\sum_{i: \lambda_i
> 2^{-n}} \lambda_i -2^{-n} = \lambda_1-2^{-n}$. As $\trn{\rho-\tilde{I}} \le \epsilon
-2^{-n}$ we conclude that $\lambda_1 \le \epsilon$. It follows that

$$S(\rho) \ge \sum_{i>1} \lambda_i \log(\lambda_i^{-1}) \ge \sum_{i>1} \lambda_i \cdot n > (1 - \epsilon)n.$$

which completes the proof.
\end{proof}

\begin{proof}(Of Lemma \ref{lem:local-change})
$f(x) = x \log x^{-1}$ is concave. Therefore, for
$\lambda_j=\delta_j+\epsilon=(1 - \frac{\epsilon}{\delta_1 -
\delta_j}) \delta_j + \frac{\epsilon}{\delta_1 - \delta_j} \delta_1$
we get: $f(\lambda_j) \ge (1 - \frac{\epsilon}{\delta_1 - \delta_j})
f(\delta_j) + \frac{\epsilon}{\delta_1 - \delta_j} f(\delta_1)$.
Similarly, $f(\lambda_1) \ge \frac{\epsilon}{\delta_1 - \delta_j}
f(\delta_j) + (1 - \frac{\epsilon}{\delta_1 - \delta_j})
f(\delta_1)$. Together, $f(\lambda_j) + f(\lambda_1) \ge f(\delta_j)
+ f(\delta_1)$. Therefore,
\begin{eqnarray*}
S(\rho) - S(\sigma) & = & \lambda_1 \log \lambda_1^{-1} + \lambda_j \log \lambda_j^{-1} - \delta_1 \log \delta_1^{-1} - \delta_j \log \delta_j^{-1} \\
& = & f(\lambda_1) + f(\lambda_j) - f(\delta_1) - f(\delta_j) \ge 0.
\end{eqnarray*}
\end{proof}

\subsection{Closure under boolean formula} \label{sec:closure}

In order to prove that $\QED$ reduces to $\QSD$ we need to
generalize another classical result about $\SZK$ to $\QSZK$, namely,
closure under boolean formula. A special case of this is, e.g., that
if $\Pi \in \QSZK$ then the promise problem that accepts $(x_1,x_2)$
if $x_1 \in \Pi_{yes}$ or $x_2 \in \Pi_{yes}$ and rejects if both
$x_i$ are in $\Pi_{no}$, is also in $\QSZK$. Notice that as we deal
with promise problems we have yes instances and no instances and
also "undefined" instances, and therefore we need to say how to
treat those "undefined" instances in our formula. We define:

\begin{deff}
\label{def:chi} For a promise problem $\Pi$, the
\emph{characteristic function} of $\Pi$ is the map $\chi_{\Pi} :
\B^* \to \set{0,1,\star}$ given by
$$ \chi_{\Pi}(x) =   \left\{%
\begin{array}{ll}
    1       & \text{if } x \in \Pi_Y \\
    0       & \text{if } x \in \Pi_N \\
    \star   & \text{otherwise}
\end{array}%
\right. $$
\end{deff}

and,

\begin{deff}
A \emph{partial assignment} to variables $v_1,\ldots,v_k$ is
$k$-tuple $\overline{a}=(a_1,\ldots,a_k) \in \set{0,1,\star}^k$. For
a propositional formula $\phi$ on variables $v_1,\ldots,v_k$ the
evaluation $\phi(\overline{a})$ is recursively defined as follows:
 \begin{center}
  \begin{tabular*}{0.75\textwidth}%
     {@{\extracolsep{\fill}}cccr}
  $v_i(\overline{a}) = a_i$, &
  $(\phi \wedge \psi)(\overline{a}) = \left\{%
\begin{array}{ll}
    1       & \text{if } \phi(\overline{a})=1 \text{ and }  \psi(\overline{a})=1 \\
    0       & \text{if } \phi(\overline{a})=0 \text{ or }  \psi(\overline{a})=0 \\
    \star   & \text{otherwise} \\
\end{array}%
\right.$  \\ \\
    $(\neg \phi)(\overline{a}) = \left\{%
\begin{array}{ll}
    1       & \text{if } \phi(\overline{a})=0 \\
    0       & \text{if } \phi(\overline{a})=1 \\
    \star   & \text{otherwise} \\
\end{array}%
\right.$   &
  $(\phi \vee \psi)(\overline{a}) = \left\{%
\begin{array}{ll}
    1       & \text{if } \phi(\overline{a})=1 \text{ or }  \psi(\overline{a})=1 \\
    0       & \text{if } \phi(\overline{a})=0 \text{ and }  \psi(\overline{a})=0 \\
    \star   & \text{otherwise} \\
\end{array}%
\right.$  \\
  \end{tabular*}
 \end{center}

\end{deff}

Notice that, e.g., $0 \wedge \star=0$ even though one of the inputs
is "undefined" in $\Pi$.

With that we define:

\begin{deff}
For any promise problem $\Pi$, we define a new promise problem
$\Phi(\Pi)$ as follows:
\begin{eqnarray*}
\Phi(\Pi)_Y & = & \set{(\phi,x_1,\ldots,x_m) : \phi(\chi_{\Pi}(x_1),\ldots,\chi_{\Pi}(x_m)) = 1} \\
\Phi(\Pi)_N & = & \set{(\phi,x_1,\ldots,x_m) :
\phi(\chi_{\Pi}(x_1),\ldots,\chi_{\Pi}(x_m)) = 0}
\end{eqnarray*}
\end{deff}

The following is an adaptation of the classical proof of \cite{SV99}
to the quantum setting:

\begin{thm}
\label{thm:formula-closure} For any promise problem $\Pi \in \QSZK$,
$\Phi(\Pi) \in \QSZK$.
\end{thm}

\begin{proof}
Let $\Pi$ be any promise problem in $\QSZK$. Since $\QSD$ is
$\QSZK$-complete, $\Pi$ reduces to $\QSD$. This induces a reduction
from $\Phi(\Pi)$ to $\Phi(\QSD)$. Thus, it suffice to show that
$\Phi(\QSD)$ reduces to $\QSD$.

\begin{claim}
$\Phi(\QSD)$ reduces to $\QSD$.
\end{claim}
\begin{proof}
Let $w=(\phi, (X_0^1, X_1^1),\ldots,(X_0^m, X_1^m))$ be an instance
of $\Phi(\QSD)$. By applying De Morgan's Laws, we may assume that
the only negations in $\phi$ are applied directly to the variables.
(Note that De Morgan's Laws still hold in our extended boolean
algebra.) By the polarization lemma (Theorem \ref{thm:polarization})
and by the closure of $\QSZK$ under complement (as was shown by
\cite{W02}), we can construct in polynomial time pairs of circuits
$(Y_0^1,Y_1^1),\ldots,(Y_0^m,Y_1^m)$ and
$(Z_0^1,Z_1^1),\ldots,(Z_0^m,Z_1^m)$ such that:

\begin{eqnarray*}
(X_0^i,X_1^i) \in \QSD_Y & \Rightarrow & \trnorm{\ket{Y_0^i} -
\ket{Y_1^i}} \ge 1 - \frac{1}{3|\phi|} \text{ and }
\trnorm{\ket{Z_0^i} - \ket{Z_1^i}} \le \frac{1}{3|\phi|} \\
(X_0^i,X_1^i) \in \QSD_N & \Rightarrow & \trnorm{\ket{Y_0^i} -
\ket{Y_1^i}} \le \frac{1}{3|\phi|} \text{ and }
\trnorm{\ket{Z_0^i} - \ket{Z_1^i}} \ge 1 - \frac{1}{3|\phi|} \\
\end{eqnarray*}

The reduction outputs the pair of circuits
$(\text{BuildCircuit}(\phi, 0), \text{BuildCircuit}(\phi, 1))$,
where BuildCircuit is the following recursive procedure:

\begin{boxit}
$\textbf{BuildCircuit}(\psi, b)$
\begin{enumerate}
    \item If $\psi = v_i$, output $Y_b^i$.
    \item if $\psi = \neg v_i$, output $Z_b^i$.
    \item If $\psi = \tau \vee \mu$, output $\text{BuildCircuit}(\tau, b) \tensor \text{BuildCircuit}(\mu, b)$.
    \item If $\psi = \tau \wedge \mu$, output $\half (\text{BuildCircuit}(\tau, 0) \tensor \text{BuildCircuit}(\mu, b)) +
                                               \half (\text{BuildCircuit}(\tau, 1) \tensor \text{BuildCircuit}(\mu, 1-b))$.
\end{enumerate}
\end{boxit}

Notice that the number of recursive calls equals the number of
sub-formula of $\phi$, and therefore the procedure runs in time
polynomial in $|\psi|$ and $|X^j_i|$, i.e., polynomial in its input
length.

We now turn to proving correctness by induction. For a sub-formula
$\tau$ of $\phi$, let

\begin{eqnarray*}
\Delta(\tau) &=& \trnorm{(\text{BuildCircuit}(\tau, 0) -
\text{BuildCircuit}(\tau, 1))\ket{0}} \end{eqnarray*}

We claim:

\begin{claim}
Let $\overline{a} =
(\chi_{\QSD}(X_0^1,X_1^1),\ldots,\chi_{\QSD}(X_0^m,X_1^m))$.
\footnote{we remind the reader that $\chi_{\QSD}(C_1,C_2)$ was
defined in Definition \ref{def:chi}.} For every sub-formula $\psi$
of $\phi$, we have:

\begin{eqnarray*}
  \psi(\overline{a}) = 1 & \Rightarrow  & \Delta(\psi) \ge 1 - \frac{|\psi|}{3|\phi|} \\
  \psi(\overline{a}) = 0 & \Rightarrow  & \Delta(\psi) \le \frac{|\psi|}{3|\phi|}
\end{eqnarray*}
\end{claim}

\begin{proof}
By induction on the sub-formula of $\phi$. It holds for atomic
sub-formula by the properties of the $Y$'s and $Z$'s.

\begin{itemize}

\item The case $\psi = \tau \vee \mu$.

If $\psi(\overline{a}) = 1$ then either $\tau(\overline{a}) = 1$ or
$\mu(\overline{a}) = 1$. W.l.o.g., say $\tau(\overline{a}) = 1$. In
this case we have for any $i \in \B$ that $\text{BuildCircuit}(\tau,
i) = \mathcal{E}\left(\text{BuildCircuit}(\psi, i)\right)$, where
$\mathcal{E}$ is the quantum operation tracing out the registers
associated with the $\mu$ sub-formula. Thus, by Fact
\ref{fact:quantum-operator-decrease-distance} and by induction,
$$ \Delta(\psi) \ge \Delta(\tau) \ge 1 - \frac{|\tau|}{3|\phi|} \ge 1 - \frac{|\psi|}{3|\phi|}.$$
If $\psi(\overline{a}) = 0$, then both $\tau(\overline{a}) =
\mu(\overline{a}) = 0$.

Using

\begin{eqnarray*}\trn{\rho_0 \tensor \rho_1 - \sigma_0
\tensor \sigma_1} &\le& \trn{\rho_0 \tensor \rho_1 - \sigma_0
\tensor \rho_1} +
\trn{\sigma_0 \tensor \rho_1 - \sigma_0 \tensor \sigma_1} \\
&=&  \trn{\rho_0 - \sigma_0} + \trn{\rho_1 - \sigma_1}.
\end{eqnarray*}

we get

$$ \Delta(\psi) \le \Delta(\tau) + \Delta(\mu) \le \frac{|\tau|}{3|\phi|} + \frac{|\mu|}{3|\phi|} \le \frac{|\psi|}{3|\phi|}.$$

\item The case $\psi = \tau \wedge \mu$.

Using

\begin{eqnarray*}
& & \trn{\half [\rho_0 \tensor \sigma_0 +  \rho_1 \tensor \sigma_1]
-
\half [\rho_0 \tensor \sigma_1 + \rho_1 \tensor \sigma_0]} \\
&=& {1 \over 2} \trnorm{(\rho_0 - \rho_1) \tensor (\sigma_0 -
\sigma_1)} ~=~ \trnorm{\rho_0 - \rho_1} \trnorm{\sigma_0 - \sigma_1}
\end{eqnarray*}

where the equalities above follow because $2\trn{X \tensor
Y}=2\trn{X}2\trn{Y} $. We get $\Delta(\psi) = \Delta(\tau) \cdot
\Delta(\mu)$.

If $\psi(\overline{a}) = 1$, then, by induction,
$$\Delta(\psi) \ge \left( 1 - \frac{|\tau|}{3|\phi|} \right) \left( 1 - \frac{|\mu|}{3|\phi|} \right) >
1 - \frac{|\tau|+|\mu|}{3|\phi|} \ge 1 - \frac{|\psi|}{3|\phi|}.$$

If $\psi(\overline{a}) = 0$, then, w.l.o.g., say $\tau(\overline{a})
= 0$. By induction
$$\Delta(\psi) = \Delta(\tau) \cdot \Delta(\mu) \le \Delta(\tau) \le \frac{|\tau|}{3|\phi|} \le \frac{|\psi|}{3|\phi|}.$$
\end{itemize}
\end{proof}

Let $A_b = \text{BuildCircuit}(\phi, b)$. By the above claim if $w
\in \Phi(\QSD)_Y$ then $\trnorm{(A-B)\ket{0}} \ge 2/3$ and if $w \in
\Phi(\QSD)_N$ then $\trnorm{(A-B)\ket{0}} \le 1/3$. Thus the claim
follows.

\end{proof}

\end{proof}

To finish the section we observe that

\begin{claim}
$\QED \le \Phi(\QEA)$, for some formula $\Phi$.
\end{claim}

\begin{proof}
Let $(Q_0,Q_1)$ be an instance of \QED. Let $\xi_i = \tensor ^6
\ket{Q_i}$. The output of the reduction is

$$\bigvee_{t=1}^{6n} \left[((\xi_0,t) \in \QEA_Y) \wedge ((\xi_1,t) \in \QEA_N)\right].$$

If $(Q_0,Q_1) \in \QED_Y$ then $S(\xi_0) \ge S(\xi_1) + 3$. Thus,
there exists an integer $t$ such that $(\xi_0,t) \in \QEA_Y$ and
$(\xi_1,t) \in \QEA_N$. On the other hand, if $(Q_0,Q_1) \in \QED_N$
then $S(\xi_1) \ge S(\xi_0) + 3$. Thus, every integer $t$ is either
greater than $S(\xi_0) + 1$ or smaller then $S(\xi_1) - 1$. That is,
for every $t$, $(\xi_0,t) \in \QEA_N$ or $(\xi_1,t) \in \QEA_Y$.
\end{proof}

In particular, the closure under formula implies that if $\QEA \le
\SD$ then $\QED=\Phi(\QEA) \le \QSD$, as desired.

\subsection{$\QSD \le \QED$} \label{sec:QSD-QED}

\subsubsection{Some quantum information backgroud}

The proof of the following facts can be found in \cite{NC00}.

\begin{fact}
\label{fact:joint-entropy-theorem} \textbf{(Joint entropy theorem)}
Suppose $p_i$ are probabilities, $\ket{i}$ are orthogonal states for
a system $A$, and $\rho_i$ is any set of density operators for
another system B. Then
$$S \left( \sum_i p_i \ketbra{i}{i} \tensor \rho_i \right) = H(p_i)+ \sum_i p_i S(\rho_i).$$
\end{fact}

\begin{fact}
\label{fact:Fannes} \textbf{(Fannes' inequality)} Suppose $\rho$ and
$\sigma$ are density matrices over a Hilbert space of dimension $d$.
Suppose further that the trace distance between them satisfies $t =
\trnorm{\rho - \sigma} \le 1/e$. Then
$$|S(\rho) - S(\sigma)| \le t( \ln d - \ln t ).$$
\end{fact}

The following lemma is taken from \cite{ANTV02}. It can be proved
using Holevo's bound.

\begin{lemm}
\label{lem:ANTV} \textbf{(Lemma 3.2, \cite{ANTV02})} Let $\rho_0$
and $\rho_1$ be two density matrices, and let $\rho = \half(\rho_0 +
\rho_1)$. If there exists is a measurement with outcome $0$ or $1$
such that making the measurement on $\rho_b$ yields the bit $b$ with
probability at least $p$, then
$$S(\rho) \ge \half[S(\rho_0) + S(\rho_1)] + (1-H(p)).$$
\end{lemm}

Combining the lemma with Fact \ref{fact:trn} we get

\begin{lemm}
\label{lem:ANTV-with-trace-norm} Let $\rho_0$ and $\rho_1$ be two
density matrices, and let $\rho = \half(\rho_0 + \rho_1)$. Then
$$S(\rho) \ge \half[S(\rho_0) + S(\rho_1)] + (1-H(\half + \frac{\trn{\rho_0 - \rho_1}}{2})).$$
\end{lemm}

\subsubsection{The proof}

\begin{thm}
For any $0 \le \alpha < \beta^2\le 1$, $\QSD_{\alpha, \beta} \le
\QED$.
\end{thm}

\begin{proof}
Given circuits $Q_0, Q_1$, We first apply the polarization lemma
(Theorem \ref{thm:polarization}) with $n=m_0$ and obtain circuits
$R_0, R_1$. We then construct two circuits $Z_0$ and $Z_1$ as
follows. $Z_1$ is implemented by a circuit which first applies a
Hadamard gate on a single qubit $b$, measures $b$ and then
conditioned on the result it applies either $R_0$ or $R_1$. The
output of $Z_1$ is $\half \ketbra{0}{0} \tensor \ket{R_0} + \half
\ketbra{1}{1} \tensor \ket{R_1}$. $Z_0$ is the same as $Z_1$ except
that $b$ is traced out. The output of $Z_0$ is $\half \ket{R_0} +
\half \ket{R_1}$. The output of $C$ is simply a qubit in the
completely mixed state.

The reduction outputs the following pair of circuits: $(Z_0 \tensor
Z_0 \tensor C, Z_1 \tensor Z_1)$.

The intuition behind the reduction is as follows. First consider the
case when $\ket{R_0}$ and $\ket{R_1}$ are very close to each other.
the matrix $\half \ket{R_0} + \half \ket{R_1}$ is very close both to
$\ket{R_0}$ and to $\ket{R_0}$, thus we "lose" the bit of
information telling us which circuit was activated. However, the
matrix $\half \ketbra{0}{0} \tensor \ket{R_0} + \half \ketbra{1}{1}
\tensor \ket{R_1}$ does contain this bit of information, i.e. has
increased entropy. On the other hand, whenever $\ket{R_0}$ and
$\ket{R_1}$ are very far, the matrix $\half \ket{R_0} + \half
\ket{R_1}$ does contain almost the same amount of information as
$\half \ketbra{0}{0} \tensor \ket{R_0} + \half \ketbra{1}{1} \tensor
\ket{R_1}$.


\begin{claim}
If $(Q_0, Q_1) \in (\QSD_{\alpha, \beta})_{NO}$ then $(Z_0 \tensor
Z_0 \tensor C, Z_1 \tensor Z_1) \in \QED_{NO}$
\end{claim}

\begin{proof}
We know that $\trnorm{\ket{Q_0} - \ket{Q_1}} \le \alpha$. By the
Polarization lemma (Theorem \ref{thm:polarization}) we get
$\trnorm{\ket{R_0} - \ket{R_1}} \le 2^{-m_0}$. By the joint-entropy
theorem (Fact \ref{fact:joint-entropy-theorem}),

$$S(\ket{Z_1}) = \half (S(\ket{R_0})+ S(\ket{R_1}))~+~1.$$

On the other hand, $\ket{Z_0}$ is very close both to $\ket{R_0}$ and
to $\ket{R_1}$. Specifically, $\trnorm{\ket{Z_0} - \ket{R_1}} =
\trnorm{\half \ket{R_0} - \half \ket{R_1}} \le 2^{-m_0}$. Therefore,
by Fennes inequality (Fact \ref{fact:Fannes}) $|S(\ket{Z_0}) -
S(\ket{R_1})| \le 2^{-m_0} \cdot \poly(m_0) \leq 0.1~$, for large
enough $m_0$. Similarly, $|S(\ket{Z_0}) - S(\ket{R_0})| \le 0.1$. It
follows that

$$|S(\ket{Z_0}) - \half (S(\ket{R_0})+ S(\ket{R_1}))| \le 0.1.$$

Combining the two equations we get $ S(\ket{Z_1}) - S(\ket{Z_0}) \ge
0.9$. Thus, $S(\ket{Z_1 \tensor Z_1}) - S(\ket{Z_0 \tensor Z_0
\tensor C}) \ge 2*0.9 - 1 = 0.8$. Therefore, $(Z_0 \tensor Z_0
\tensor C, Z_1 \tensor Z_1) \in \QED_{NO}$
\end{proof}

\begin{claim}
If $(Q_0, Q_1) \in (\QSD_{\alpha, \beta})_{YES}$ then $(Z_0 \tensor
Z_0 \tensor C, Z_1 \tensor Z_1) \in \QED_{YES}$
\end{claim}

\begin{proof}
By the Polarization lemma (Theorem \ref{thm:polarization})
$\trnorm{\rho_0 - \rho_1} \ge 1 - 2^{-m_0}$. Using the Holevo bound
(Lemma \ref{lem:ANTV-with-trace-norm}) we get that $S(\ket{Z_0}) \ge
\half[S(\rho_0) + S(\rho_1)] + 1-H(\half + \frac{\trn{\rho_0 -
\rho_1}}{2}) \ge \half[S(\rho_0) + S(\rho_1)] + 1 - H(2^{-m_0})$. By
Fact \ref{fact:joint-entropy-theorem} we know that $S(\ket{Z_1} =
\half (S(\rho_0) + S(\rho_1)) + 1$. Therefore, $S(\ket{Z_1}) -
S(\ket{Z_0}) = H(2^{-m_0}) < 0.1$ for sufficiently large $m_0$.

In particular, $S(\ket{Z_1 \tensor Z_1}) - S(\ket{Z_0 \tensor Z_0
\tensor C}) \le 2*0.1 - 1 = -0.8$ and $(Z_0 \tensor Z_0 \tensor C,
Z_1 \tensor Z_1) \in \QED_{YES}$

\end{proof} 
\end{proof} 

\section{Quantum extractors}
\label{sec:app:quantum-extractors}

\begin{lemm}
If $T:L(V) \to L(V)$ is a $(D=2^d,\lbar)$ quantum expander, then for
every $t>0$, $T$ is also a $(k=n-t,d,\epsilon)$ quantum extractor
with $\epsilon=2^{t/2} \cdot \lbar$.
\end{lemm}

\begin{proof}
$T$ has a dimension $1$ eigenspace $W_1$ with eigenvalue $1$,
spanned by the norm $1$ eigenvector $v_1={1 \over \sqrt{N}}I$ (where
$dim(V)=N$). Our input $\rho$ is a density matrix and therefore $\la
\rho ~|~ v_1 \ra = {1 \over \sqrt{N}} \Tr(\rho) = {1 \over
\sqrt{N}}$. In particular $\rho -{1 \over \sqrt{N}}v_1=\rho-\nI$ is
perpendicular to $W_1$. Therefore,

\begin{eqnarray*}
||T(\rho)-\nI||_2^2 &=&||T(\rho-\nI)||_2^2  ~ \le ~ \lbar^2
||\rho-\nI||_2^2 \\
&=& \lbar^2 [ ||\rho||_2^2-\la \rho | \nI \ra - \la \nI | \rho \ra +
||\nI||_2^2] = \lbar^2 [ ||\rho||_2^2 -{1 \over N}] \le \lbar^2
||\rho||_2^2.
\end{eqnarray*}

Plugging $H_2(\rho) \ge \minentropy(\rho) \ge k=n-t$ we see that
$||T(\rho)-\nI||_2^2 \le \lbar^2 2^{-(n-t)}$. Using Cauchy-Schwartz

\begin{eqnarray*}
\trn{T(\rho)-\nI} & \le & \sqrt{N} ||T(\rho)-\nI||_2 ~\le~  \sqrt{N}
\cdot \lbar \cdot 2^{-{n-t \over 2}} ~=~ 2^{t/2} \cdot \lbar ~=~
\epsilon.
\end{eqnarray*}
\end{proof}

\begin{thma}
Any $(D,\lbar)$ quantum expander satisfies $\lbar \ge \frac{2}{3
\sqrt{3D}}$.
\end{thma}

\begin{proof}
Let $E$ be a $(D,\lbar)$ quantum expander operating on the space of
$n$ qubits. Let $d = \log D$, and let $\delta>0$ be a constant to be
fixed later. We first apply Lemma \ref{lem:expander-extractor} with
$t=d-2\log\delta$ to deduce that $E$ is a
$(n-d+2\log\delta,d,\delta^{-1} 2^{d/2} \lbar)$ quantum extractor.

The proof idea is to take a density matrix which is uniform on a set
of "small size". Applying the extractor yields a density matrix
close to the completely mixed state. Such a matrix must have a high
rank. On the other hand, because we started with a low-rank matrix,
the resulting density matrix cannot have a too-high rank (since $E$
is $D$-regular).

Formally, let $\rho \in D(V)$ be a density matrix of a flat
(classical) probability distribution over a set of size
$2^{n-d+2\log\delta}$. By definition, $\minentropy(\rho) =
n-d+2\log\delta$. Also,  $\log (\rank(\rho)) = n-d+2\log\delta$.

Using the quantum extractor definition we get that $E(\rho)$ is
$\delta^{-1} 2^{d/2} \lbar$-close to the completely mixed state.
Hence,
$$\rank(E(\rho)) \ge (1- \delta^{-1} 2^{d/2} \lbar)2^n.$$

On the other hand, since $E$ is $2^d$-regular, $E(\rho)$ is a sum of
$2^d$ matrices. Each of these matrices has rank
$2^{n-d+2\log\delta}$. Hence,
$$\rank(E(\rho)) \le 2^d 2^{n-d+2\log\delta} = 2^n \delta^2.$$

Combining the two inequalities gives
$$\lbar \ge \delta(1 - \delta^2)2^{-d/2} = \frac{\delta(1 - \delta^2)}{\sqrt{D}}.$$
Taking $\delta = 1/\sqrt{3}$ completes the proof.
\end{proof}

\end{document}